\documentclass[useAMS,usenatbib]{mn2e}
\usepackage{graphicx}
\usepackage{color}
\usepackage{soul} 
\usepackage{url}
\usepackage{txfonts}
\usepackage{longtable}

\definecolor{myred}{rgb}{0.7,0.0,0.2}

\definecolor{myblue}{rgb}{0.0,0.2,0.7}

\definecolor{mygreen}{rgb}{0.2,0.7,0.0}

\title[Symbiotic stars in M33]{A Survey of the Local Group of Galaxies for Symbiotic Binary Stars. I. First detection of symbiotic stars in M33}
\author[J. Miko{\l}ajewska et al.]{Joanna Miko{\l}ajewska$^{1}$\thanks{E-mail: mikolaj@camk.edu.pl}, Michael M. Shara$^{2}$,  Nelson Caldwell$^{3}$, 
\newauthor Krystian I{\l}kiewicz$^{1}$ and David Zurek$^{2}$ 
\\
$^{1}$Nicolaus Copernicus Astronomical Center, Polish Academy of Sciences, Bartycka 18, PL 00--716 Warsaw, Poland\\
$^{2}$Department of Astrophysics, American Museum of Natural History, Central Park West at 79th Street, New York, NY 10024, USA\\
$^{3}$Harvard-Smithsonian Center for Astrophysics, Cambridge, MA 02138, USA\\
}

\begin{document}

\date{Accepted  Received }

\pagerange{\pageref{firstpage}--\pageref{lastpage}} \pubyear{}

\maketitle

\label{firstpage}

\begin{abstract}
We present and discuss initial selection criteria and first results in M33
from a systematic search for extragalactic symbiotic stars. 
We show that the presence of diffuse interstellar gas emission 
can significantly contaminate the spectra of symbiotic star candidates. This important effect forces upon us a more stringent working definition of an extragalactic symbiotic star.
We report the first detections and spectroscopic characterisation of 12 symbiotic binaries in M33. 
We found that four of our systems contain carbon-rich giants. In another two of them the giant seems to be a Zr-enhanced MS star, while the remaining six objects host M-type giants. The high number ratio of C to M giants in these binaries is consistent with the low metallicity of M33.
The spatial and radial velocity distributions of these new symbiotic binaries are consistent with a wide range of progenitor star ages.
\end{abstract}



\section{Introduction}

Symbiotic stars (SySt) are interacting binaries, in which an evolved giant (either a normal red giant (RG) in S-type, or a Mira surrounded by an opaque dust shell in D-type SySt) tranfers mass to a hot, luminous and compact companion which is usually a white dwarf (WD). The interacting stars are embedded in rich and complex surroundings, including both ionized and neutral regions, accretion/excretion  disks, interacting winds, and jets. SySt are important and luminous tracers of the late evolutionary phases of low- and medium-mass binary stars, and excellent laboratories to test models of close binary star evolution. They should be detectable throughout the Milky Way. 
Moreover, the rapid mass transfer in these WD + giant binaries suggests that some of them might be single-degenerate (SD) progenitors of type Ia supernovae (SNIa). Many SySt will evolve to become double WDs, and mergers of such objects are another pathway to SNIa.
The most recent extensive review of SySt is in \citet{mik2012}.

While about 300 Galactic symbiotics are known (e.g. \citealt{Bel2000};  \citealt{MMU2013} \citealt{MM2014}; \citealt{rf2014}, and references therein), and a few dozen are relatively well studied, their distances (and hence their component luminosities and other distance-related physical parameters) are poorly determined. This makes confrontation of theoretical models of SySt with the observed parameters of real stars, to test theories of their interactions and evolution, very challenging.

Fortunately, a few dozen bright SySt have been detected in the Magellanic Clouds 
(\citealt{Bel2000};  \citealt{MMU2014}, and references therein)
and recently even in members of the Local Group of galaxies 
(\citealt{Goncalves2008}; \citealt{kniazev2009}; \citealt{MCS14}, hereafter MCS14). These discoveries have been, however, in almost all cases, serendipitous. A systematic search for SySt in multiple galaxies is essential to provide samples that are large enough to be suitable for statistical analyses and tests of binary evolution theory.

The motivation and the basic concepts for this survey have already been presented by MCS14. In essence, our goal is to obtain large, complete, luminosity-limited samples of extragalactic SySt to determine their total numbers and their spatial distributions in galaxies of different types. These are strong constraints on binary stellar evolution and the progenitor masses. Large samples, all at the same distance, will enable us to determine the values and distributions of these stars' luminosity-related physical parameters.

This paper focusses on our initial selection criteria and results in M33. Our most important result is that while extragalactic SySt candidates can be easier to select than their Galactic counterparts, there is an isidious systematic effect at work that must be understood and accounted for before their confirmation and characterization is valid. This effect - the superposition of diffuse interstellar gas (DIG) in extragalactic SySt spectra - is so ubiquitous and important that we devote part of this first paper to demonstrating its importance for all future studies of extragalactic SySt. The working definition of an extragalactic symbiotic star must take into account the possible presence of DIG emission if we are to produce samples uncontaminated by imposters that are not really SySt. In this paper we present the first 12 symbiotic stars to be detected in M33, and in the subsequent paper in this series we will report on over 100 new detections in M31. As expected, the spatial distributions of these stars offer important clues to their progenitor masses and ages.

In Section~\ref{obs} we describe the selection method, and the data and their reductions for our initial M33 survey.  The coordinates, observed and dereddened spectra, and classification of the new stars we observed are presented in Section~\ref{class}. We characterize our new SySt in Section~\ref{analysis}, and
briefly summarize our results in Section~\ref{concl}.

\section{Selection method and follow-up spectroscopy}\label{obs}

\begin{figure}
\centerline{\includegraphics[width=0.99\columnwidth]{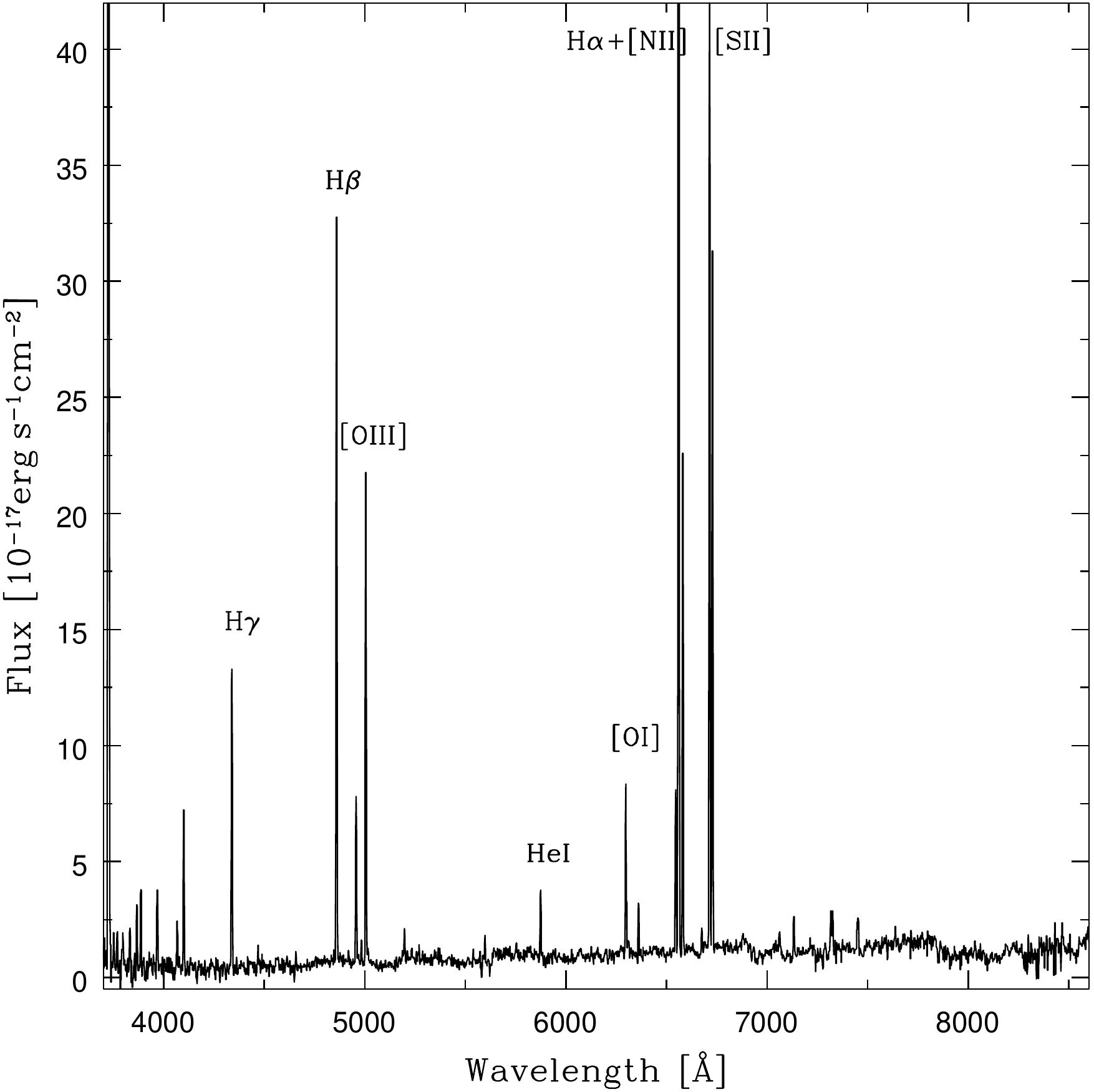}}
\centerline{\includegraphics[width=0.99\columnwidth]{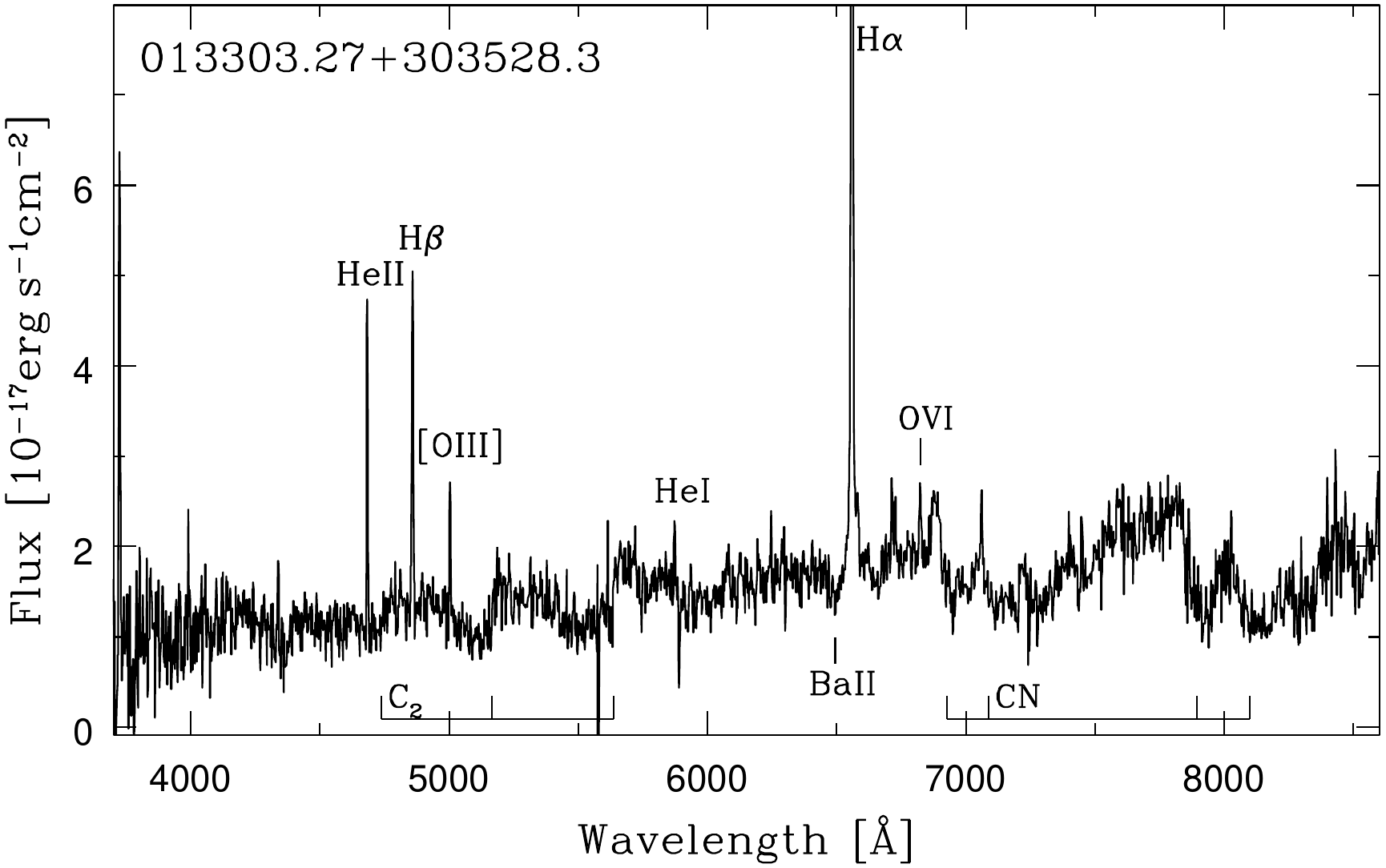}}
\caption{Spectra of two SySt candidates in M33, both showing strong absorption features of a carbon giant and strong emission lines. However, only the spectrum on the bottom can be attributed to a SySt according to our criteria. The emission lines in the top spectrum are most likely from DIG (see text).}\label{spectra1}
\end{figure}

\begin{figure}
\centerline{\includegraphics[width=0.99\columnwidth]{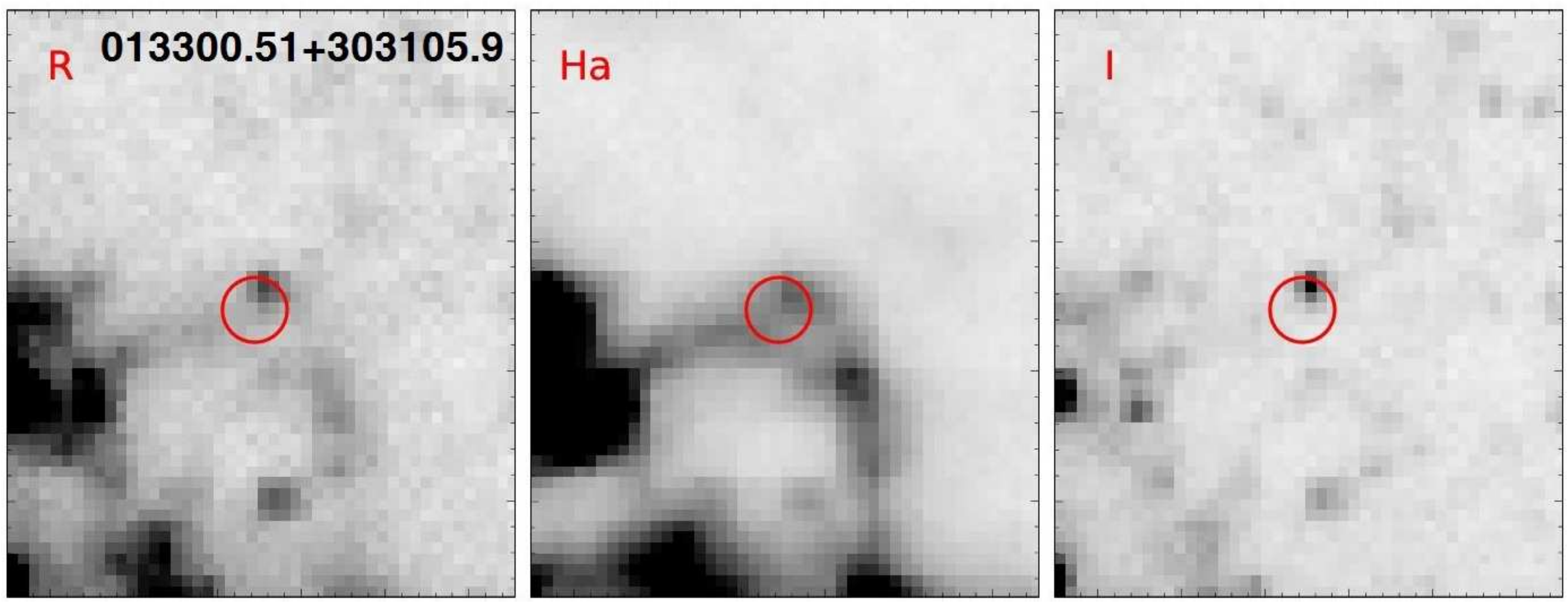}}
\centerline{\includegraphics[width=0.99\columnwidth]{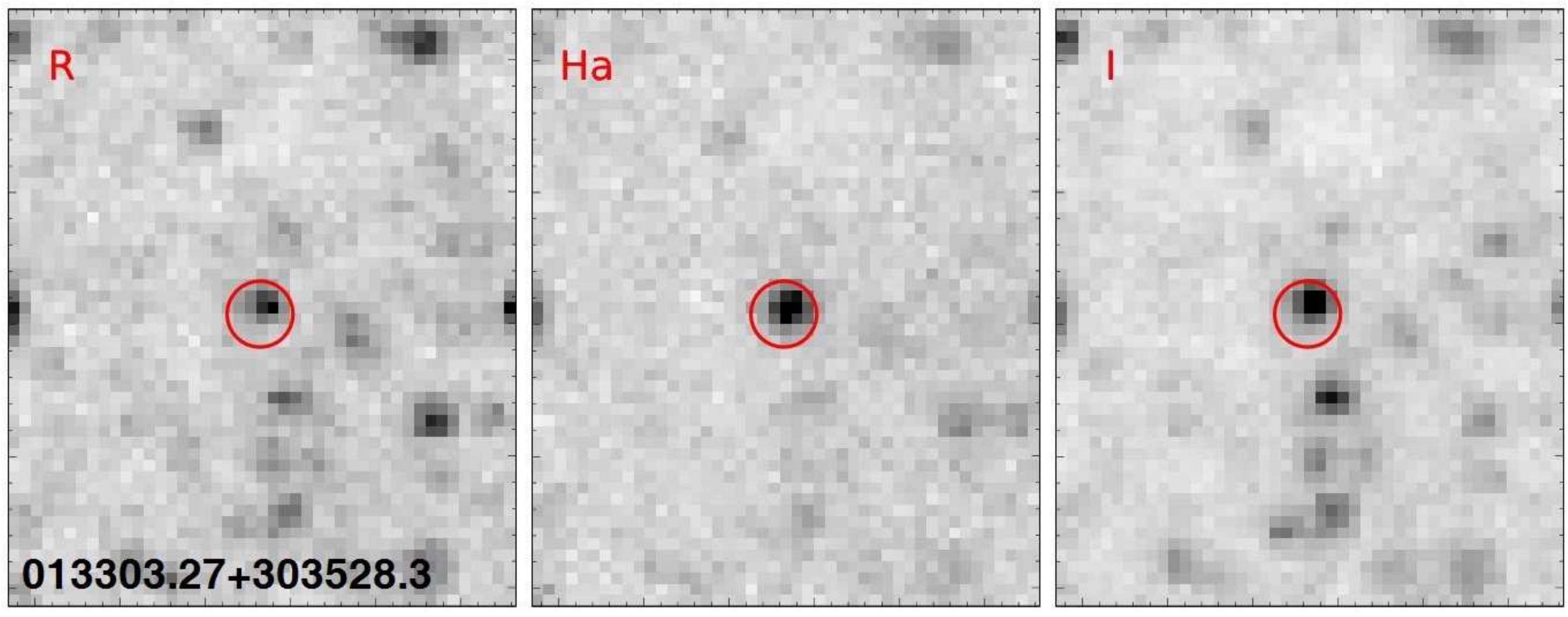}}
\caption{LGGS $R$, H$\alpha$ and $I$ images of two SySt candidates in M33, the same as those in Fig.~\ref{spectra1}.
The FoV is 16$\times$18 arcsec. The red circles have 2 arcsec diameter, and they are centered at the coordinates adopted for our Hectospec observations. The genuine SySt, M33SyS\,J013303.27+303528.3 (bottom) appears point-like in the H$\alpha$ filter while the 'false' SySt, M33\,J013300.51+303105.9 (top) is embedded in strong extended DIG emission.
 (see text).}\label{image1}
\end{figure}

We selected candidates on the basis of photometric measurements from the publicly available Local Group Galaxy Survey (LGGS) images \citep{LGGSHa}. First, we  downloaded images of M33 from the Lowell Observatory's website, and rereduced them using the aperture photometry routines in DAOPHOT \citep{stetson}.  We then created catalogs of SySt candidates with the following criteria:
(i) each star must be detected in the $VRI$  and H$\alpha$ filter images; (ii) each candidate must be red, i.e. $V-I \ga 1$, and (iii) each candidate must display Balmer emission, so that H$\alpha$-$R \la 0$. The resulting catalogs contained over 20\,000 objects in M33.

The rationale for our criteria are, of course, that SySt are simultaneously red from the presence of a cool giant and show strong emission lines from the nebula that is ionized by the hot WD. For our first observing run (September 2014 - see below) we selected those candidates with $V-I \ga 1$, $U-B \la 0$, and 
$-1 \ga {\rm H}\alpha-R \ga -4$, motivated by the $UBVI$H$\alpha$ magnitudes of SySt from MCS14. These criteria yielded  $\sim 3700$ candidates in M33.

To characterize these objects, we obtained spectra of 199  of them using the Hectospec multi-fiber positioner and spectrograph on the 6.5m MMT telescope \citep{fab05}. The Hectospec 270 gpm grating was used and provided spectral coverage from roughly $3700-9200${\AA} at a resolution of $\sim5${\AA}. The observations were made on the nights of 23 and 25 September and were reduced in the uniform manner outlined in \citet{cal09}. The frames were first de-biased and flat fielded. Individual spectra were then  extracted and wavelength calibrated. Sky subtraction is achieved with Hectospec by
averaging spectra from "blank sky" fibers from the same exposures or by offsetting the telescope by a few arcseconds. 
Standard star spectra obtained intermittently were used for flux calibration and instrumental response. These relative flux corrections were carefully applied to ensure that the relative line flux ratios would be accurate. The total exposure time was 5400\,s for all spectra.

\begin{figure*}
\centerline{\includegraphics[width=150mm]{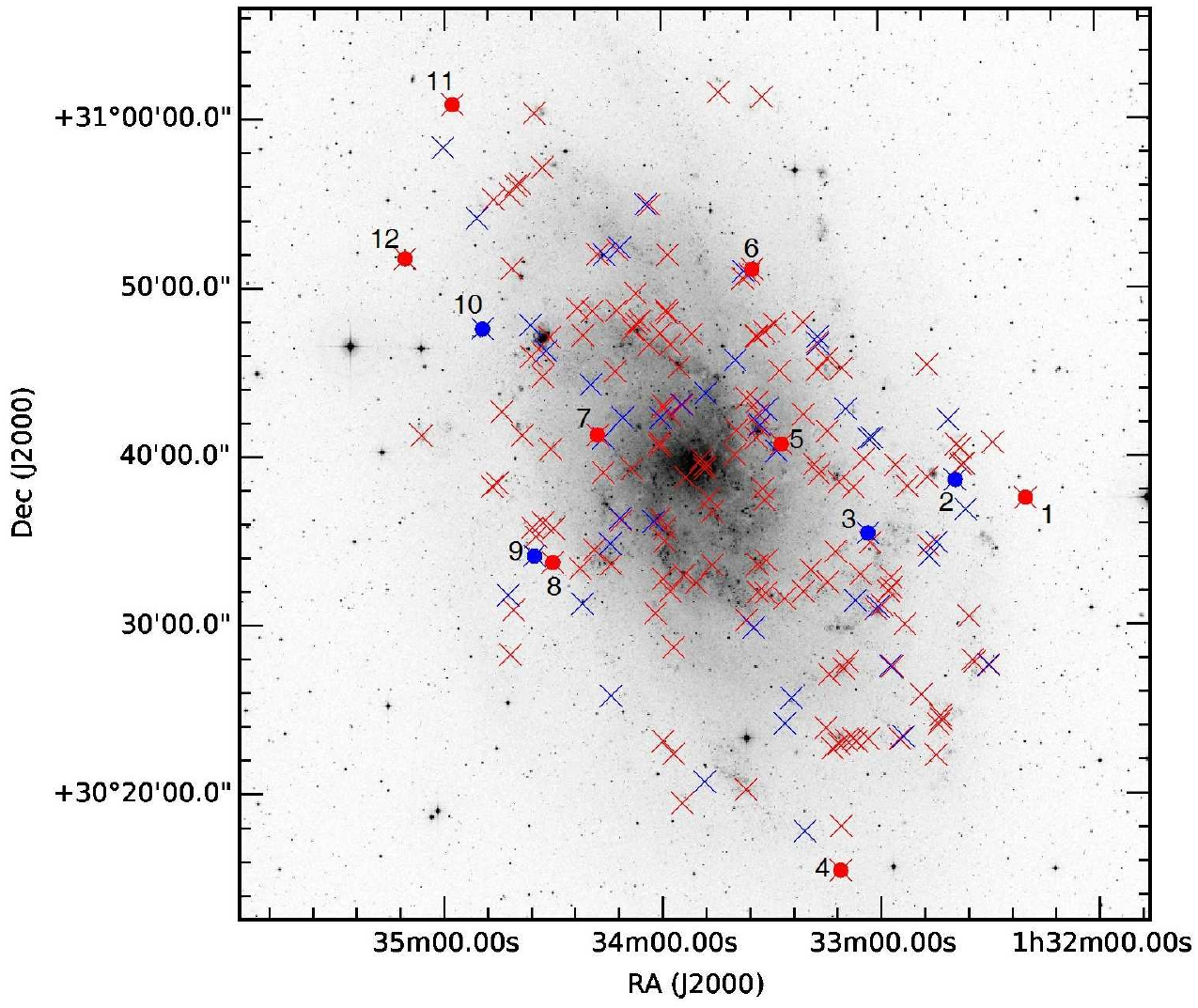}}
\caption{Distribution of all stars observed in M33 with confirmed RG: M-type (red crosses) and carbon (blue crosses). Closed circles represent the new SySt, identified by their ordinal numbers in Table~\ref{Tid}.}\label{map}
\end{figure*}

Whereas practically all objects revealed at least some H$\alpha$ emission, only 64 of them revealed the presence of a RG. Thus for the future observing runs we selected only very red candidates with $V-I \ga 2$. We also abandoned  attempts to observe objects with $I \ga 21$ (which corresponds to the faintest SySt detected in M31 by MCS14). 
That left $\sim 700$ SySt candidates, of which 283 were observed during our second observing run with Hectospec on the nights 16, 17, 19 and 26 November 2014. During this run, we also repeated observations of 17 objects (SySt and a few other interesting objects). 

\section{Identification and classification of S\lowercase{y}S\lowercase{t} in M33}\label{class}

To classify an object as a SySt, most authors adopt Kenyon's (1986)  definition, which, in addition to the presence of absorption features of a late-type giant, requires strong \mbox{H\,{\sc i}} and \mbox{He\,{\sc i}} emission lines, and additional lines with an ionizational potential of at least 30 eV (e.g. \mbox{[O\,\sc{ iii}]}) and an equivalent width $ > 1\,\rm \AA$. This is entirely appropriate for Galactic SySt, as contamination from interstellar emission lines is not observed.

The spectra of $\sim 40\,\%$ of the observed SySt candidates in M33 revealed the presence of a RG and strong emission lines satisfying the above definition (Fig.~\ref{spectra1}). However, in most of those stars, the highest IP lines are those from \mbox{[O\,\sc{ iii}]}\,N$_{1,2}$ lines. Moreover, {\it the forbidden} 
\mbox{[O\,\sc{ ii}]}, \mbox{[O\,\sc{ iii}]}, \mbox{[N\,\sc{ ii}]} {\it and} \mbox{[S\,\sc{ ii}]} {\it line ratios are consistent with low-density ($n_{\rm e} \sim 100\, \rm cm^{-3}$ or less) formation regions, and these lines presumably originate in diffuse ionized gas (DIG) in M33 rather than in the much denser symbiotic nebula}. Such DIG is present in all Local Group Galaxy disks, and is particularly abundant in star forming regions (e.g. \citealt{hp2003}), and it may significantly pollute the SySt candidate spectra. This is because in extragalactic systems our spectrographs capture the light of large volumes of space containing large quantities of line emitting, hot gas that surrounds RGs that are not part of symbiotic binaries. Fig.~\ref{spectra1} show an exemplary spectrum of a true SySt in M33, as well as that of a SySt imposter -- a RG with strong, superposed DIG lines. The SySt, M33SyS\,J013303.27+303528.3,  is clearly visible as a strong, point-like source in the H$\alpha$ image whereas the imposter, M33\,J013300.51+303105.9, is embedded in strong extended DIG emission (Fig.~\ref{image1}).

A map of all of our candidates which displayed RG with emission lines is shown in Fig.~\ref{map}.

\begin{figure}
\centerline{\includegraphics[width=0.9\columnwidth]{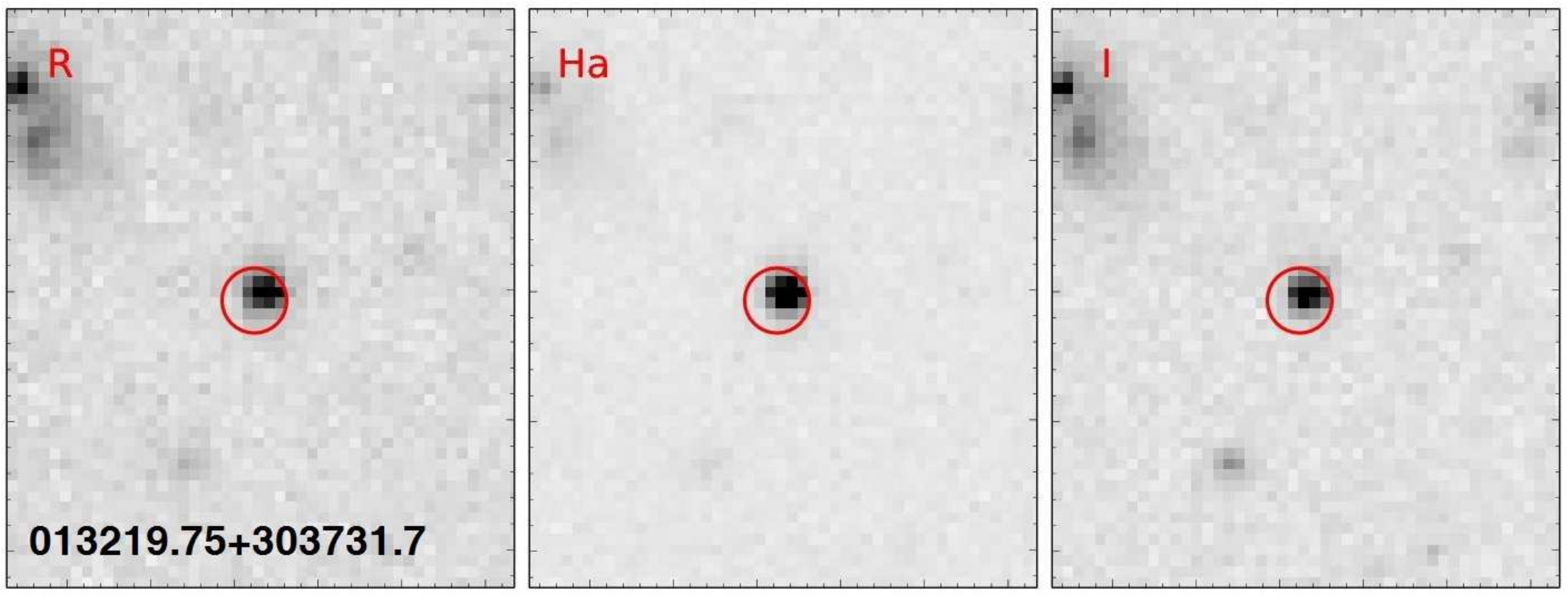}}
\centerline{\includegraphics[width=0.9\columnwidth]{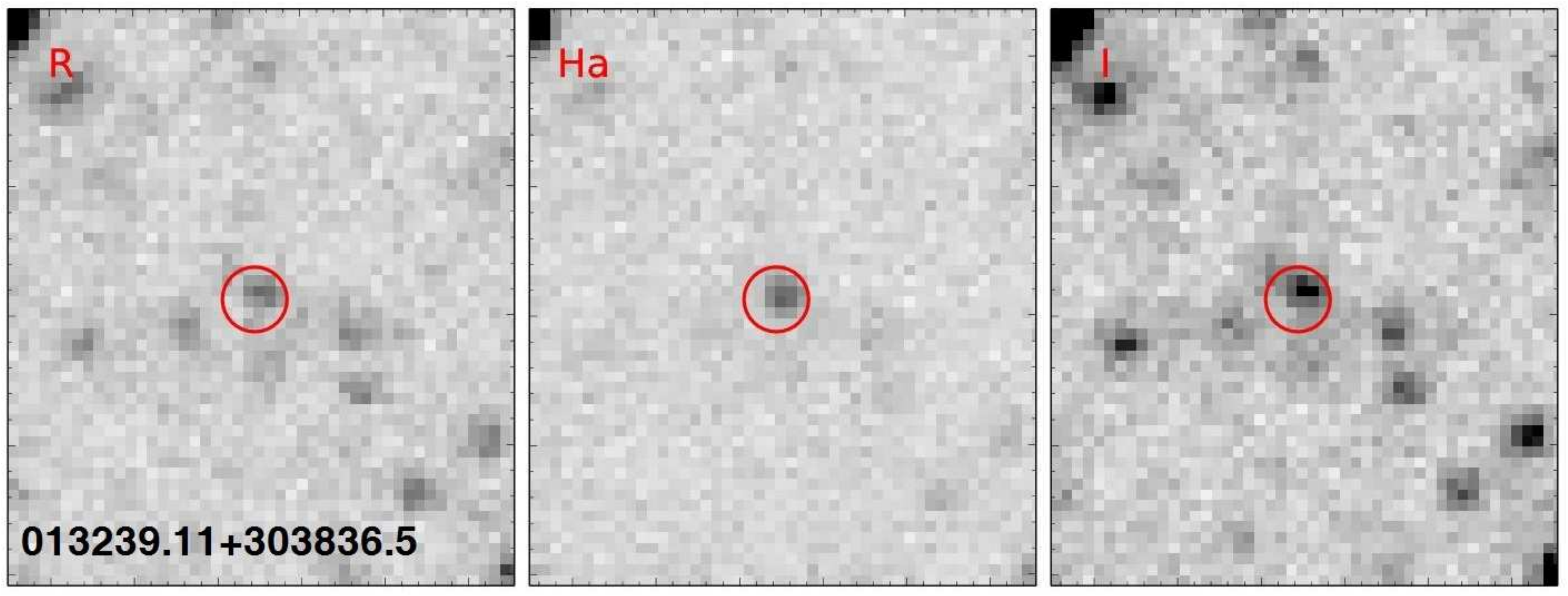}}
\centerline{\includegraphics[width=0.9\columnwidth]{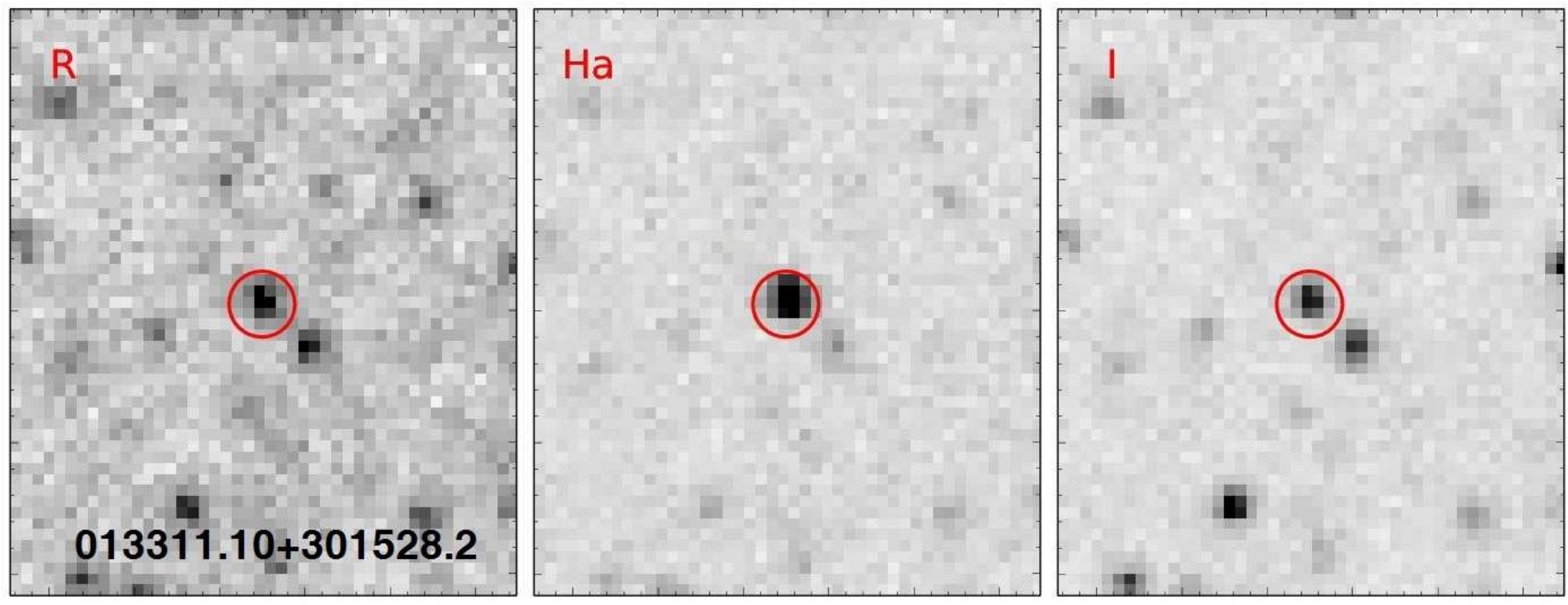}}
\centerline{\includegraphics[width=0.9\columnwidth]{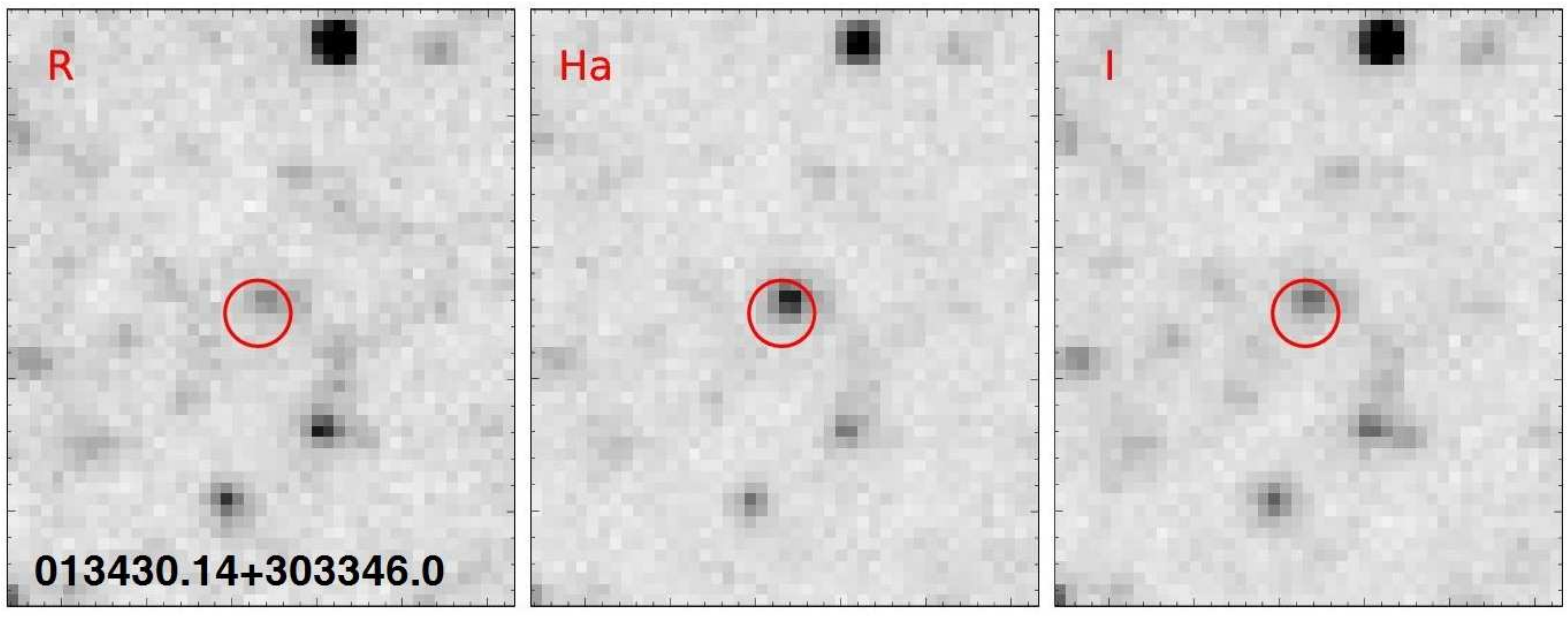}}
\centerline{\includegraphics[width=0.9\columnwidth]{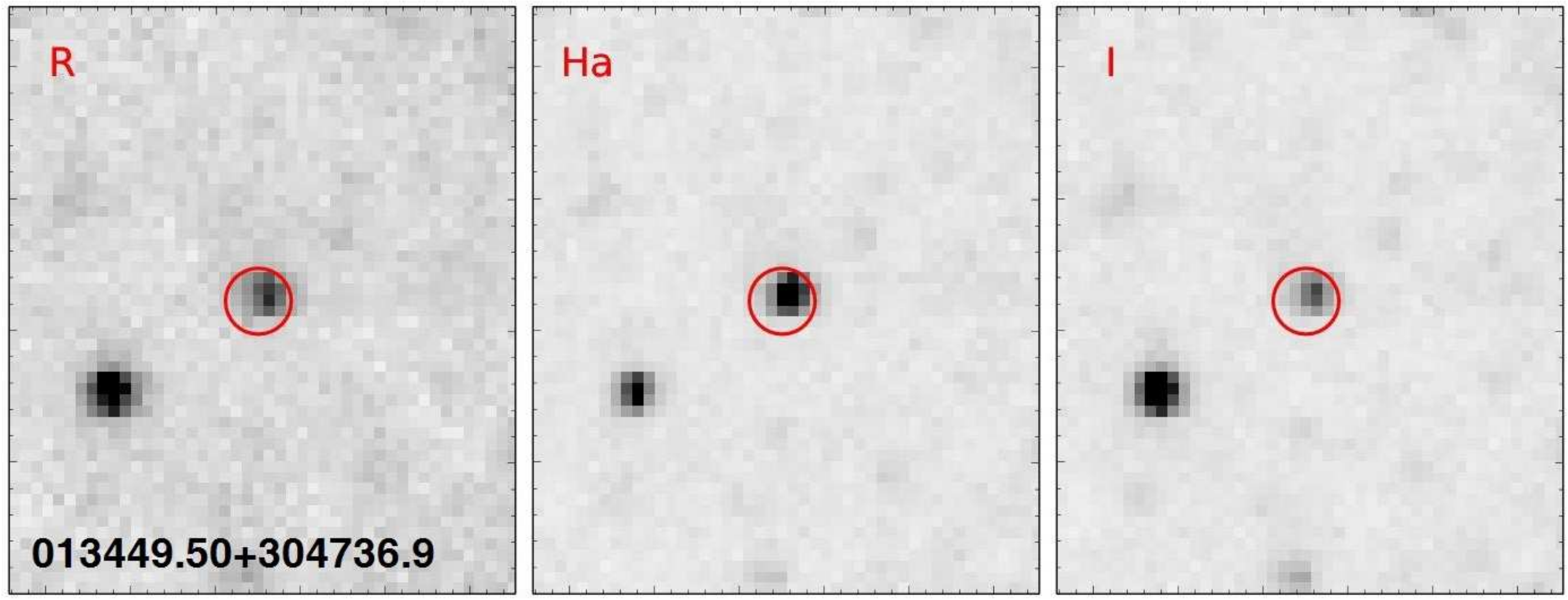}}
\centerline{\includegraphics[width=0.9\columnwidth]{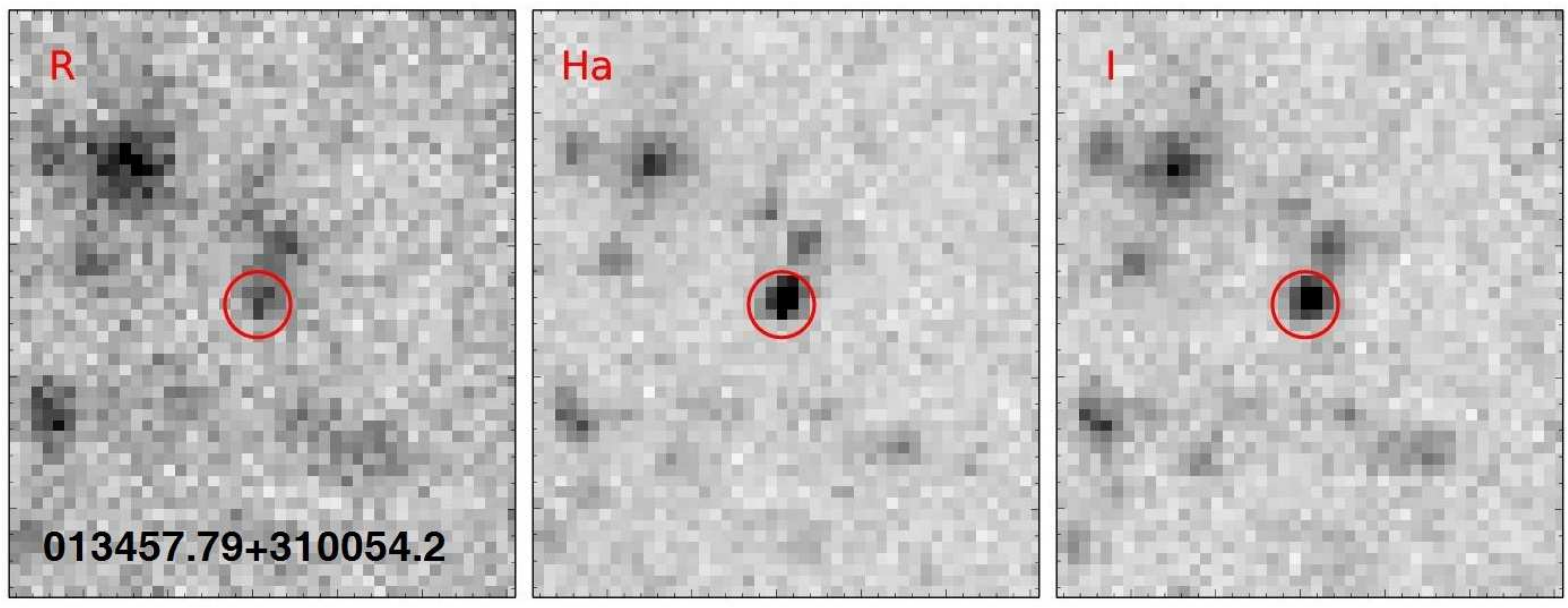}}
\centerline{\includegraphics[width=0.9\columnwidth]{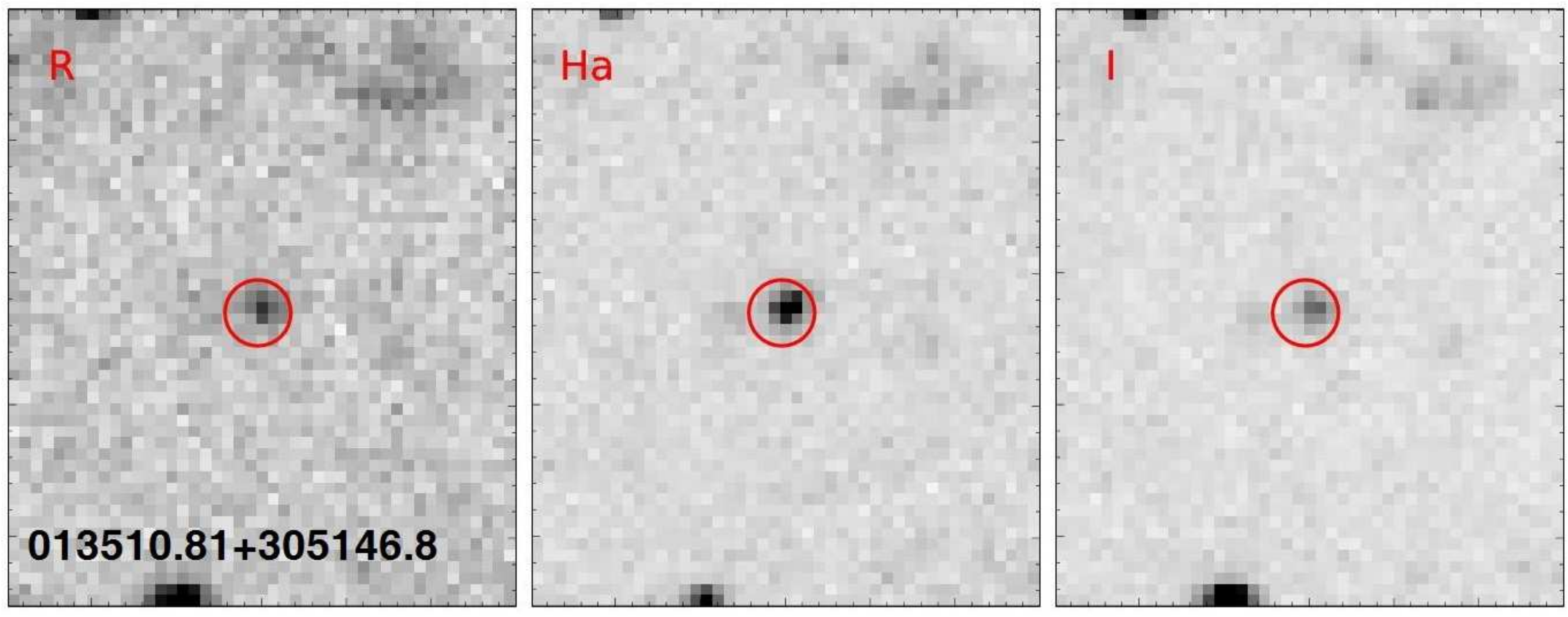}}
\caption{LGGS $R$, H$\alpha$ and $I$ images of the new SySt in M33.
The FoV is 16$\times$18 arcsec . The red circles have 2 arcsec diameter, and they are centered at the coordinates adopted for our Hectospec observations. }\label{fc1}
\end{figure}

\begin{figure}
\centerline{\includegraphics[width=0.9\columnwidth]{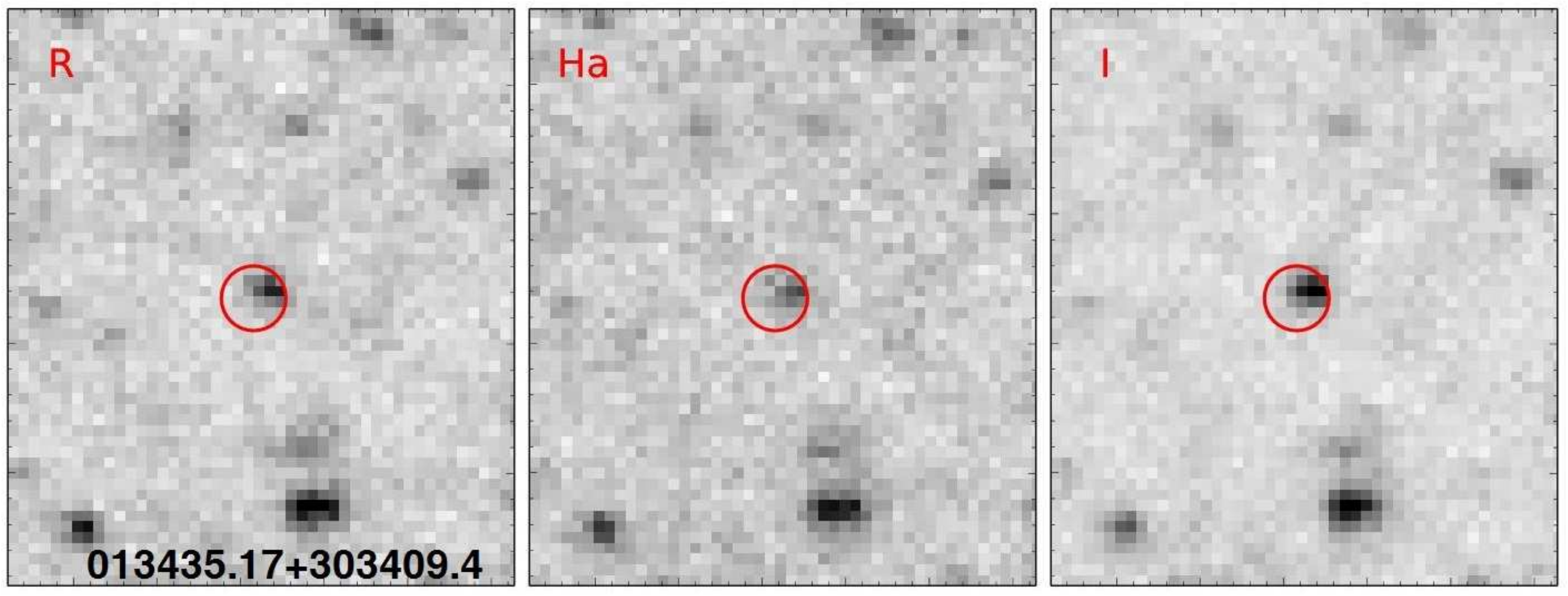}}
\caption{LGGS $R$, H$\alpha$ and $I$ images of M33SyS\,J013435.17+303409.4.
The star does not show up as an H$\alpha$-bright object.}\label{fc2}
\end{figure}

\begin{figure}
\centerline{\includegraphics[width=0.9\columnwidth]{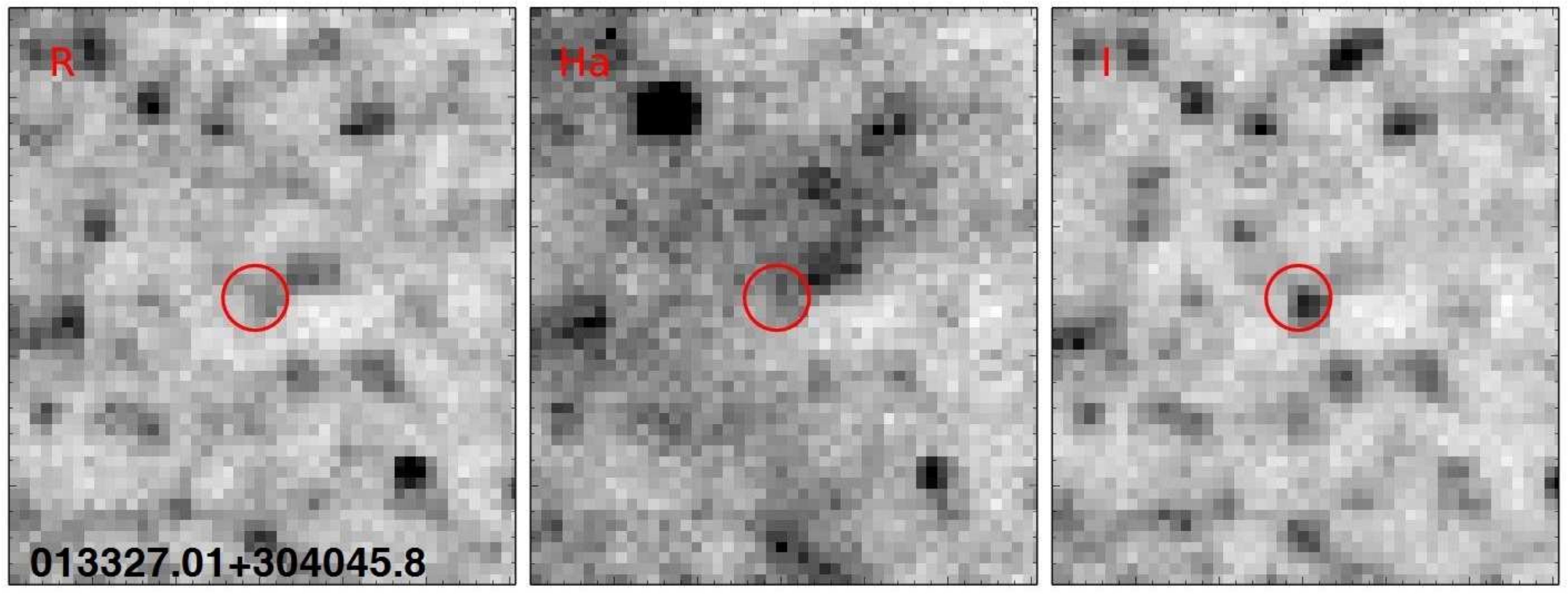}}
\centerline{\includegraphics[width=0.9\columnwidth]{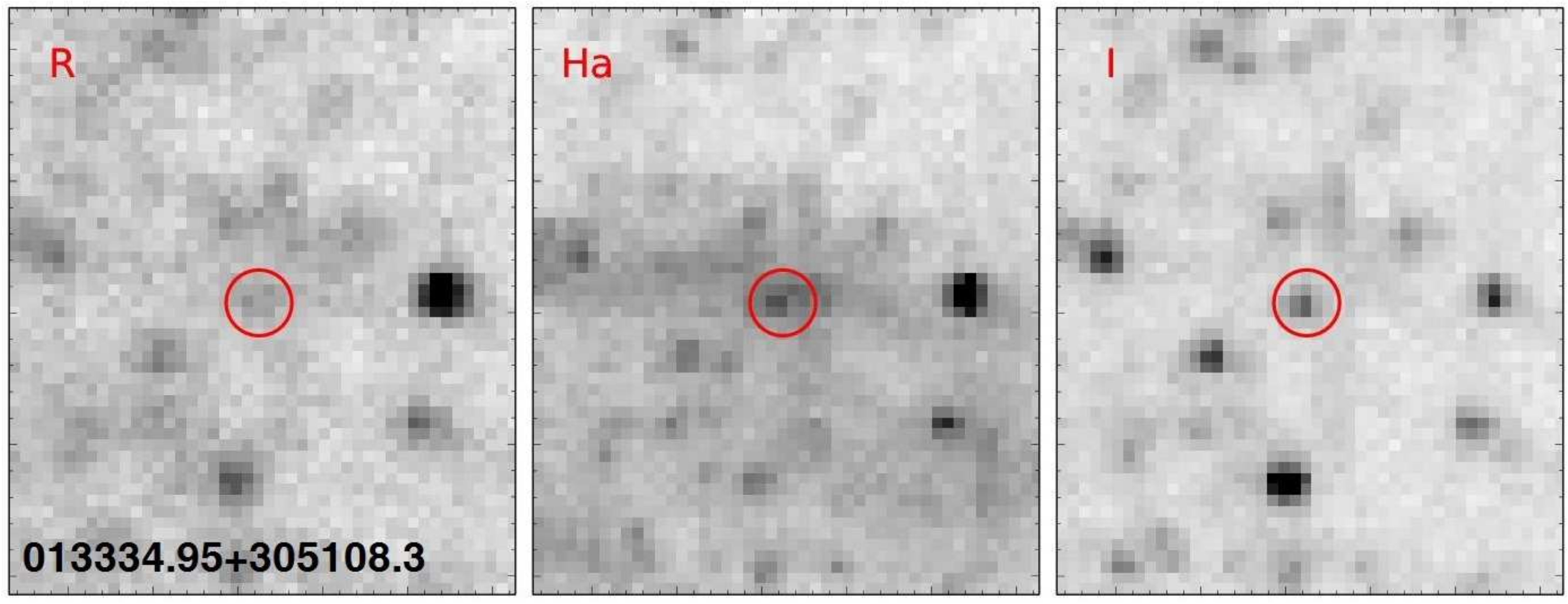}}
\centerline{\includegraphics[width=0.9\columnwidth]{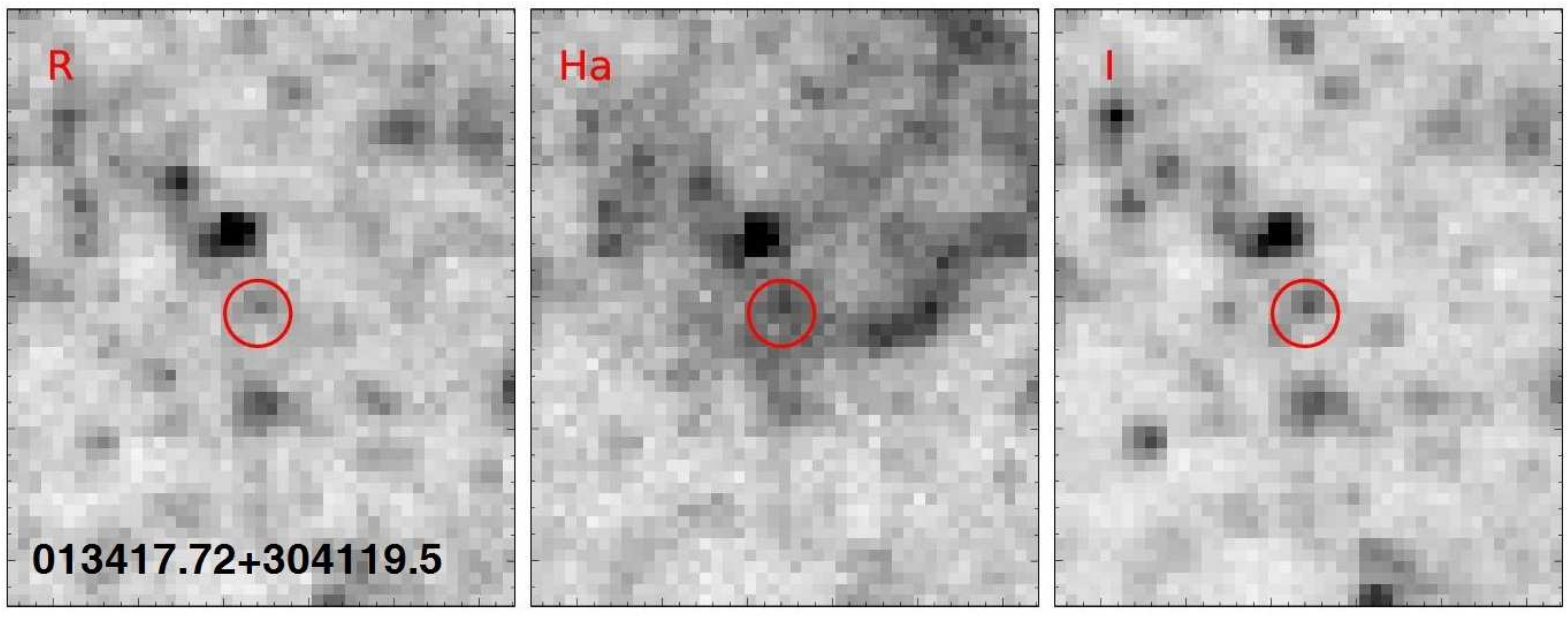}}
\caption{LGGS $R$, H$\alpha$ and $I$ images of the new SySt in M33 with strong DIG component. They appear point-like in the H$\alpha$ filter but the strong extended DIG emission is also visible (see text).}\label{fc3}
\end{figure}

\begin{table*}
 \centering
  \caption{List of new symbiotic stars with accurate coordinates, their other cross identifiers, and the reference to the number of figure presenting the spectrum.}\label{Tid}
  \begin{tabular}{@{}rllllc@{}}
  \hline
No.  &    M33SyS\,J  &  RA(2000)  &  DEC(2000)  &  Cross IDs  &  Spectrum \\
\hline
1   & 013219.75+303731.7 & 01:32:19.75 & 30:37:31.7 & LGGS\,J013219.75+303731.7 & Fig.~\ref{sp_359130}\\
     &    &  &  & [HBS2006]260101 &\\  
      &    &  &  & SDSS9\,J013219.72+30:37:31.5 & \\
     &    &  &  & SSTM3307\,J013219.72+30:37:31.5 & \\                         
2  & 013239.11+303836.5  & 01:32:39.11 & 30:38:36.5 & & Fig.~\ref{spectraC}\\	
3  & 013303.27+303528.3  & 01:33:03.27 & 30:35:28.3 &  LGGS\,J013303.27+303528.3 & Fig.~\ref{spectra1}\\
    &    &  &  & [HBS2006] & \\
     &    &  &  & SSTM3307\,J013303.24+303528.0 & \\	
4  & 013311.10+301528.2  & 01:33:11.10 & 30:15:28.2 &  [HBS2006]330166 & Fig.~\ref{spectraM}\\
      &    &  &  & SDSS9\,J013311.03+301527.7 & \\
     &    &  &  & SSTM3307\,J013311.04+301527.6 & \\  	
5  & 013327.01+304045.8  & 01:33:27.01 & 30:40:45.8 &  [HBS2006]151442 & Fig.~\ref{spectra3}\\	
6  & 013334.95+305108.3  & 01:33:34.95 & 30:51:08.3 & & Fig.~\ref{spectra3}\\	
7 & 013417.72+304119.5  & 01:34:17.72 & 30:41:19.5 & & Fig.~\ref{spectra3} \\	
8  & 013430.14+303346.0  & 01:34:30.14 & 30:33:46.0 & SST3307\,J013430.12+303345.8 & Fig.~\ref{sp_91161} \\
     &    &  &  & M33\,V0479 & \\ 	
9  & 013435.17+303409.4  & 01:34:35.17 & 30:34:09.4 &  [HBS2006]222817 & Fig.~\ref{spectraC} \\
     &    &  &  & SSTM3307\,J013435.13+303409.4 & \\	
10  & 013449.50+304736.9  & 01:34:49.50 & 30:47:36.9 &  [HBS2006]120596 & Fig.~\ref{spectraC} \\
     &    &  &  & SSTM3307\,J013449.49+304737.1 & \\	
11  & 013457.79+310054.2  & 01:34:57.79 & 31:00:54.2 & [HBS2006]30376 & Fig.~\ref{spectraM}\\
    &    &  &  & SSTM3307\,J013457.76+310054.4 & \\	
12 & 013510.81+305146.8  & 01:35:10.81 & 30:51:46.8 & [HBS2006]110047 & Fig.~\ref{spectraM}\\
     &    &  &  & SDSS9\,J013510.79+305146.5  & \\	
\hline
\end{tabular}
\begin{list}{}{}
\item LGGS... - identifier in \citet{LGGSHa}
\item {[HBS2006]} XXXXXX - number in \citet{hartman2006}
\item  SDSS9... - identifier in The SDSS Photometric Catalog Release 9 \citep{sdss9}
\item SSTM3307... - identifier in \citet{mcquinn}
\item M33\,V.... - identifier in General Catalogue of Variable Stars (Samus et al. 2007-2013)
\end{list}
\end{table*}

\begin{table*}
 \centering
  \caption{Photometric data of new SySt.}\label{Tphot}
  \begin{tabular}{@{}lcccccccccccc@{}}
  \hline
  M33SyS\,J  & \multicolumn{6}{c}{LGGS$^{1}$}  & \multicolumn{3}{c}{SDSS/CFHT}  & \multicolumn{3}{c}{2MASS$^{5}$}\\
   & $I$ & $V$--$I$ & $B$--$V$ & $U$--$B$  & $R$--$I$ & H$\alpha$--$R$ & $i$ & $g$--$r$ & $r$--$i$ & $J$ & $H$ & $K$ \\
\hline
 013219.75+303731.7 & 19.64 & 1.48 & 0.45 & -0.64 & 0.65 & -1.29 & 20.09$^{2}$ & 0.73$^{2}$ & 0.34$^{2}$ & \cr
                                   &       &     &    &   &  &   & 19.89$^{3}$ & 0.87$^{3}$ & 0.25$^{3}$ & \cr
 013239.11+303836.5 & 20.74 & 2.13 & 0.73 & 0.04 & 0.85 & -1.08 & \cr
 013303.27+303528.3 & 19.58 & 2.54 & 1.67 & -0.27 & 1.16 & -1.24 & 20.22$^{2}$ & 1.66$^{2}$ & 0.67$^{2}$ & \cr
013311.10+301528.2 & 20.32 & 2.93 & &   & 1.21 & -1.60 & 20.83$^{2}$ & 1.51$^{2}$ & 1.11$^{2}$ & 18.73 & 17.79 & 17.48 \cr
                    &       &     &    &   &  &   & 21.05$^{3}$ & 1.39$^{3}$ & 1.02$^{3}$ & \cr
 013327.01+304045.8 & 20.59 & 2.72 &  & & 1.54 & -1.06 & 21.12$^{2}$ & 1.08$^{2}$ & 1.78$^{2}$ \cr
 013334.95+305108.3 & 20.44 & 3.05 & & & 1.24 & -1.24 & \cr
 013417.72+304119.5 & 20.33 & 2.84 & & & 1.24 & -1.20 \cr
 013430.14+303346.0 & 20.17 & 2.48 & 0.48 & -0.65 & 1.15 & -1.31 \cr
013435.17+303409.4 & 20.59 & 3.20 & & & 1.55 & 0: & 20.55$^{2}$ & 1.72$^{2}$ & 0.83$^{2}$ & 17.97 & 17.02 & 16.74 \cr
 013449.50+304736.9 &19.46 & 2.02 & 1.71 & -0.33 & 0.84 & -1.09 & 20.02$^{2}$ & 1.61$^{2}$ & 0.53$^{2}$ \cr
 013457.79+310054.2 & 20.54 & 2.98 & & & 1.41 & -1.06 & 20.93$^{2}$ & 1.24$^{2}$ & 0.97$^{2}$ & 18.90 & 17.96 & 17.92 \cr
 013510.81+305146.8 & 20.29 & 2.18 & 0.82 & -0.63 & 0.85 & -1.51 & 20.98$^{2}$ & 1.30$^{2}$ & 0.79$^{2}$ \cr
                    &       &     &    &   &  &   & 21.28$^{4}$ & 1.91$^{4}$ & 0.61$^{4}$ & \cr
\hline
\end{tabular}
\begin{list}{}{}
\item $^{1}$ Our measurements on the LGGS images
\item $^{2}$ The average values from the CFHT photometry \cite{hartman2006}
\item $^{3}$ SDSS photometry on JD 2455123
\item $^{4}$ SDSS photometry on JD 2454861
\item $^{5}$ \citet{cioni2008}
\end{list}
\end{table*}

In light of the discussion above, we propose that stronger criteria must be adopted to accept an extragalactic star as a SySt. {\it In addition to RG features and strong} \mbox{H\,{\sc i}} {\it lines, the presence of} \mbox{He\,{\sc ii}}\,4686 {\it and higher ionization emission lines excited by a source with} $T \ga 10^5$\,K {\it must be present.} This is the definition originally proposed by \citet{Allen1984}. These much stronger criteria leave us with only a dozen SySt in M33, which are presented in this paper. The point, of course, is to NOT accept as a SySt any RG that displays emission lines from DIG. Such objects are SySt imposters. Counting them as real SySt will greatly inflate our SySt population statistics, and skew our perception of true SySt spatial distributions towards star regions rich in ionized hydrogen.

We emphasize that many well-known Galactic SySt do NOT show a detectable optical \mbox{He\,{\sc ii}}\,4686 emission line.
To quantify this issue, we estimated the  \mbox{He\,{\sc ii}}\,4686/H$\beta$ intensity ratio from the 146 spectra of 110 SySt collected by \citet{mz2002}.  Adopting an optimistic detection limit for our survey, \mbox{He\,{\sc ii}}\,4686/H$\beta \ga 0.1$, we find that 46 out of the 146 spectra of known SySt - about 1/3 - would not satisfy Allen's SySt definition. A more conservative detection limit, namely \mbox{He\,{\sc ii}}\,4686/H$\beta \ga 0.3$ will detect only about half of known Galactic SySt. Moreover, in the case of a SySt with a C-rich giant, a weak \mbox{He\,{\sc ii}}\,4686 line can be also easily buried in the C$_2$\,4737 band. 
This effect is clearly visible in the spectra of the known SySt LMC S63 taken near the hot component eclipse \citep{ilkiew2015}.
We maintain, however, that this is a price that the SySt community must be willing to pay to prevent wholesale contamination of extragalactic SySt catalogs with imposters that have nothing to do with SySt.

We note that with high enough S/N spectra, additional criteria can be applied to differentiate SySt from DIG-dominated RG. These e.g. involve the \mbox{[O\,{\sc iii}]} and \mbox{He\,{\sc i}} diagnostic diagrams, which cleanly distinguish between dense SySt nebulae and lower density planetary nebulae and \mbox{H\,{\sc ii}} regions (for details see MCS14, and references therein). 
Additionally, RG with broad H$\alpha$ ($\rm FW \ga 500\,  km\,s^{-1}$) and without the low-density DIG (\mbox{[S\,{\sc ii}]}, \mbox{[N\,{\sc ii}]} and \mbox{[O\,{\sc ii}]}) are very likely low IP SySt, similar to the known Galactic SySt  CH Cyg and the quiescent symbiotic recurrent novae T CrB and RS Oph which show only Balmer emission most of the time.
However, in this study we concentrate only on those {\it unambiguous} SySt with detectable \mbox{He\,{\sc ii}}.
\begin{figure}
\centerline{\includegraphics[width=0.99\columnwidth]{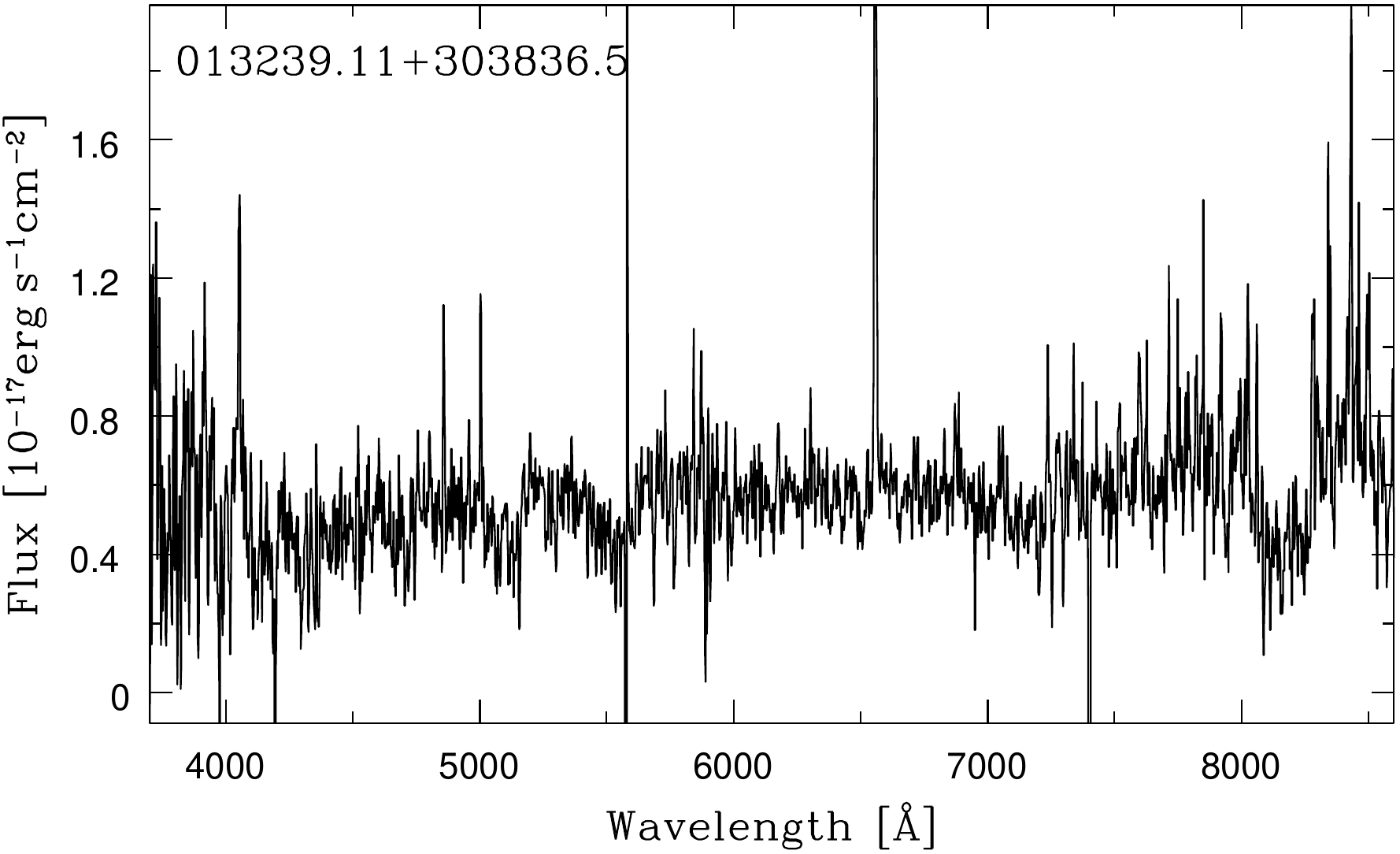}}
\centerline{\includegraphics[width=0.99\columnwidth]{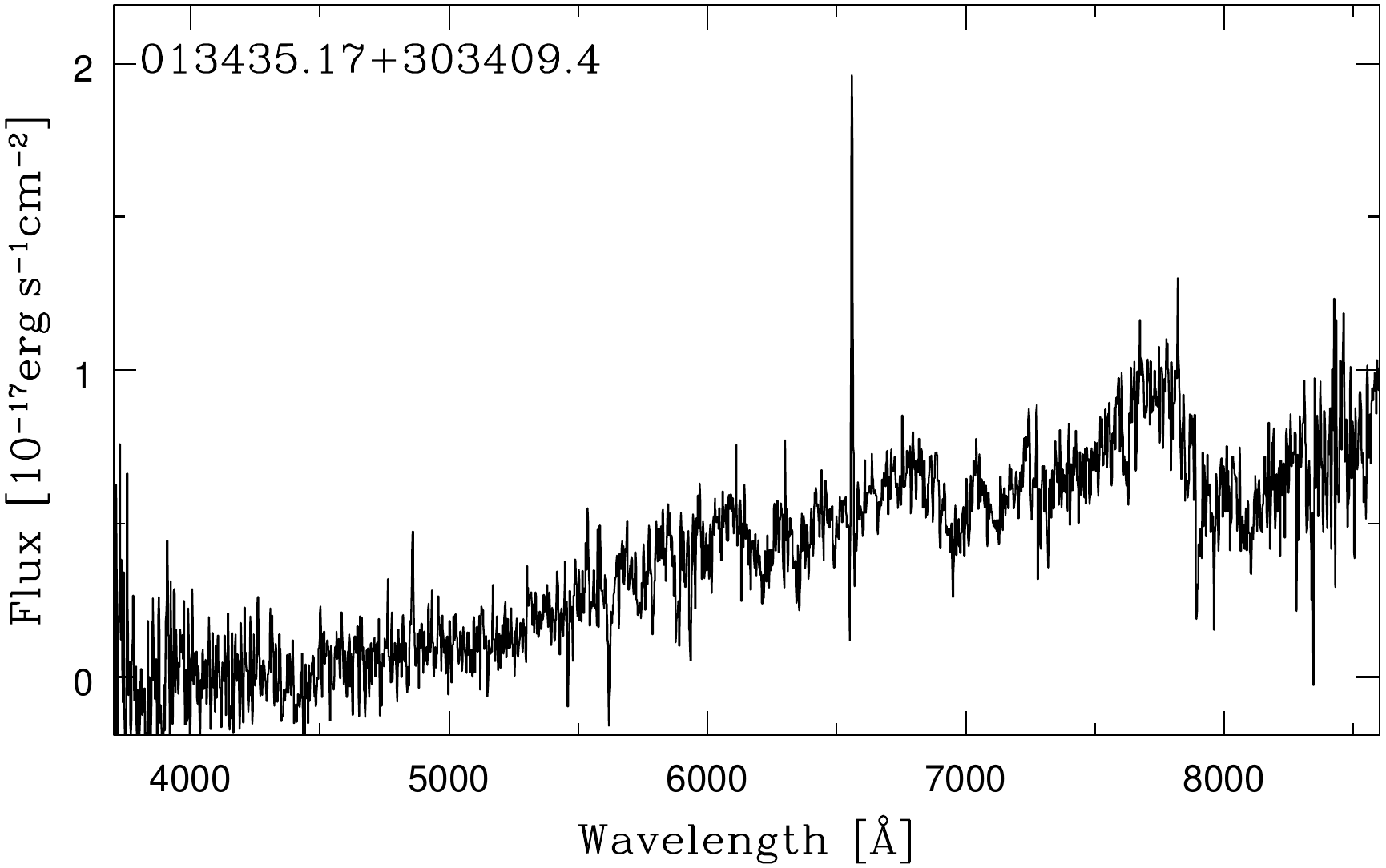}}
\centerline{\includegraphics[width=0.99\columnwidth]{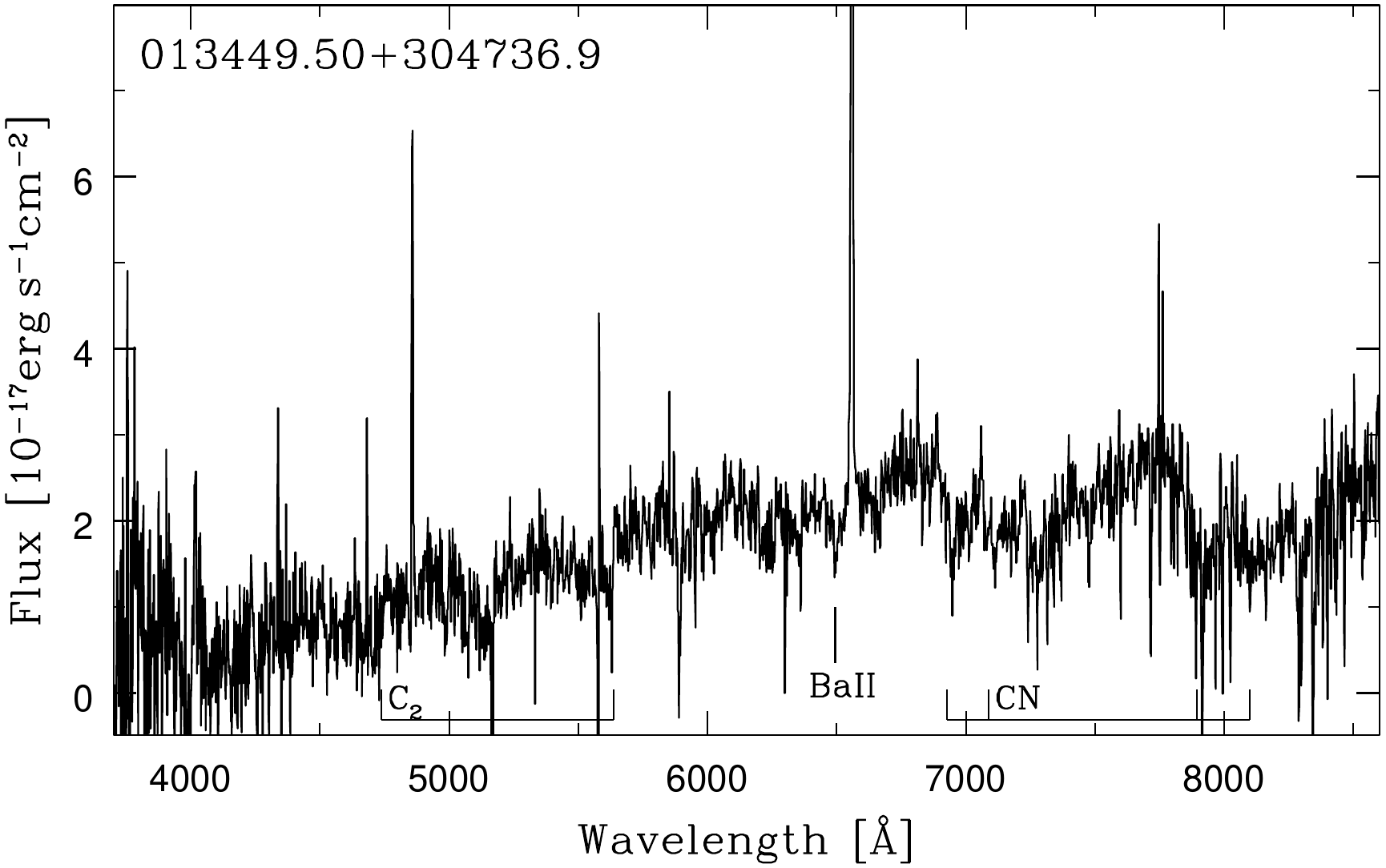}}
\caption{Spectra of SySt in M33 with C giants. The strongest C$_2$ and CN absorption bands are marked on the bottom spectrum.}\label{spectraC}
\end{figure}

\begin{figure}
\centerline{\includegraphics[width=0.99\columnwidth]{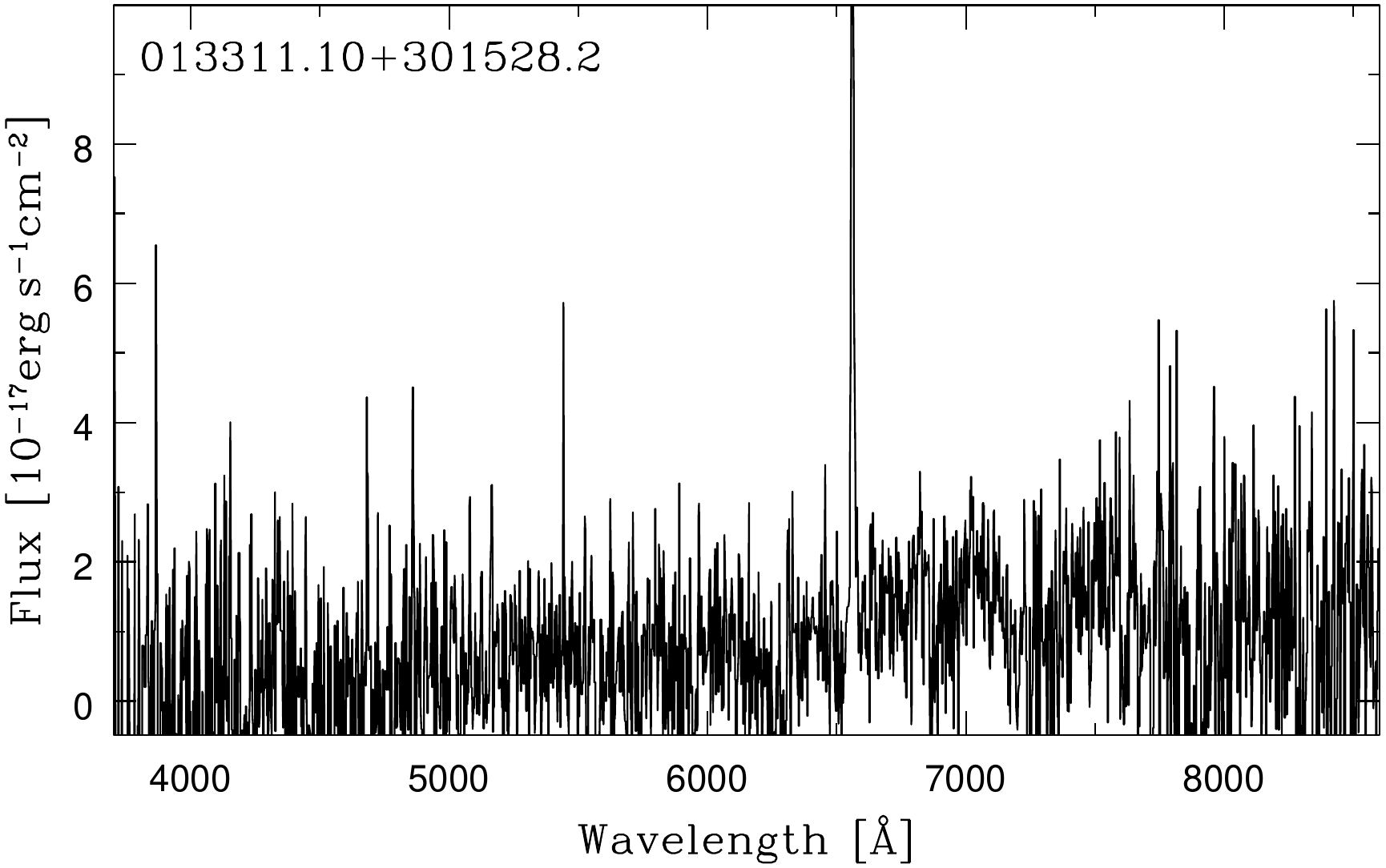}}
\centerline{\includegraphics[width=0.99\columnwidth]{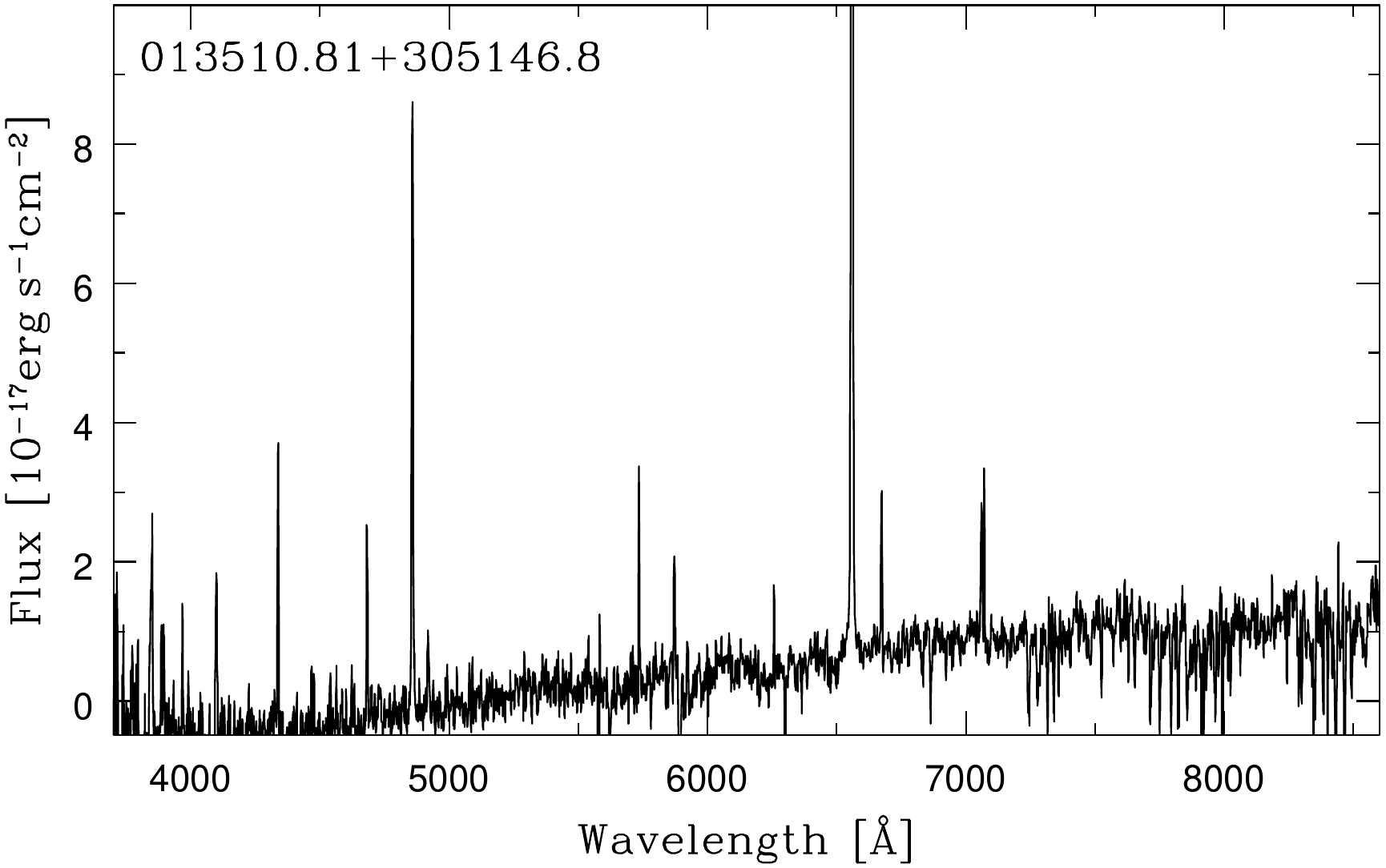}}
\centerline{\includegraphics[width=0.99\columnwidth]{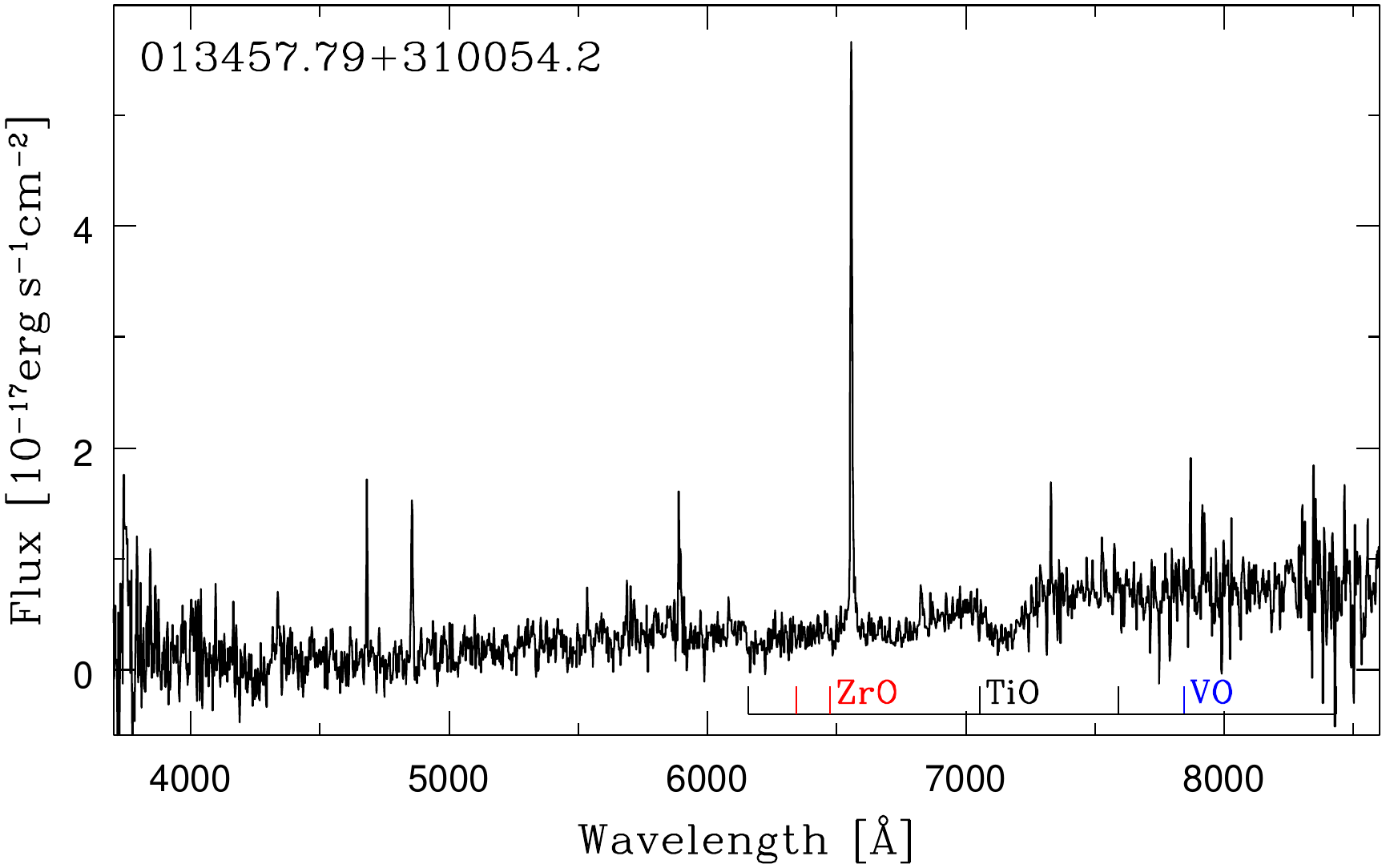}}
\caption{Spectra of SySt in M33 with M/MS giants. The strongest TiO$_2$, VO and ZrO absorption bands are marked on the bottom spectrum.}\label{spectraM}
\end{figure}

Table~\ref{Tid} lists the names and coordinates of the new SySt. Their locations in M33 are shown on the map in  Fig.~\ref{map}, and their finder charts in Figs.~\ref{image1}--\ref{fc3}.
As in MCS14, each object's name was built from its position, which resulted in the format M33SyS\,JHHMMSS.ss+DDMMSS.s where the acronym M33SyS indicates that they are also SySt.
We used VizieR\footnote{http://vizier.u-strasbg.fr/viz-bin/VizieR} to search  for matches to our new SySt in all catalogs available in the CDS database.  
The resulting matches within 0.5 arcsec radius are listed in Table~\ref{Tid}. In particular, only two of our new SySt were included in the LGGS catalogue (\citealt{massey2006}, \citealt{LGGSHa}). 

The $UBVRI$H$\alpha$ mag measured on the LGGS images, and the  $JHK$ and $gri$ (Sloan filters) photometry collected from the literature are given in Table~\ref{Tphot}. 
Table~\ref{Tid} also provides the reference number of the figure presenting the spectrum normalized to $I$ magnitudes. The accuracy of the flux calibration depends on the variability of our objects which is discussed in Section~\ref{variability}.
The emission line fluxes and other measurements from the spectra are listed in Tables~\ref{Tsp} --~\ref{Tpar}.

The absorption features of a cool component are present in all dozen spectrographically confirmed SySt in our sample.
In particular, four SySt show strong CN bands, and three of them also strong C$_2$ bands, indicating carbon rich (C) red giants (Fig.~\ref{spectra1}, and Fig.~\ref{spectraC}). All of them are N-type carbon stars except M33SyS\,J013239.11+303836.5 which is of CH type as indicated by the presence of a strong G-band of CH at 4300\,\AA. 
The spectrum of M33SyS\,J013435.17+303409.4, in addition to strong CN bands, also shows strong \mbox{Ba\,{\sc ii}} 6495 line, and it is very similar to the symbiotic C-rich Mira H1-45 \citep{MMU2013}. It is also the coolest objects in our sample, and its light curve (Section~\ref{variability}) suggests that it is indeed a Mira variable.
In the remaining objects the TiO bands are present, indicative of 
M-type giants (Fig.~\ref{spectraM}\, --~\ref{sp_91161}). 
Additionally, M33SyS\,J013510.81+305146.8 and M33SyS\,J013430.14+303346.0 show clearly detectable ZrO bands which indicates some s-process enrichment; we note this by classifying their red components as 'MxS'.

\begin{figure}
\centerline{\includegraphics[width=0.99\columnwidth]{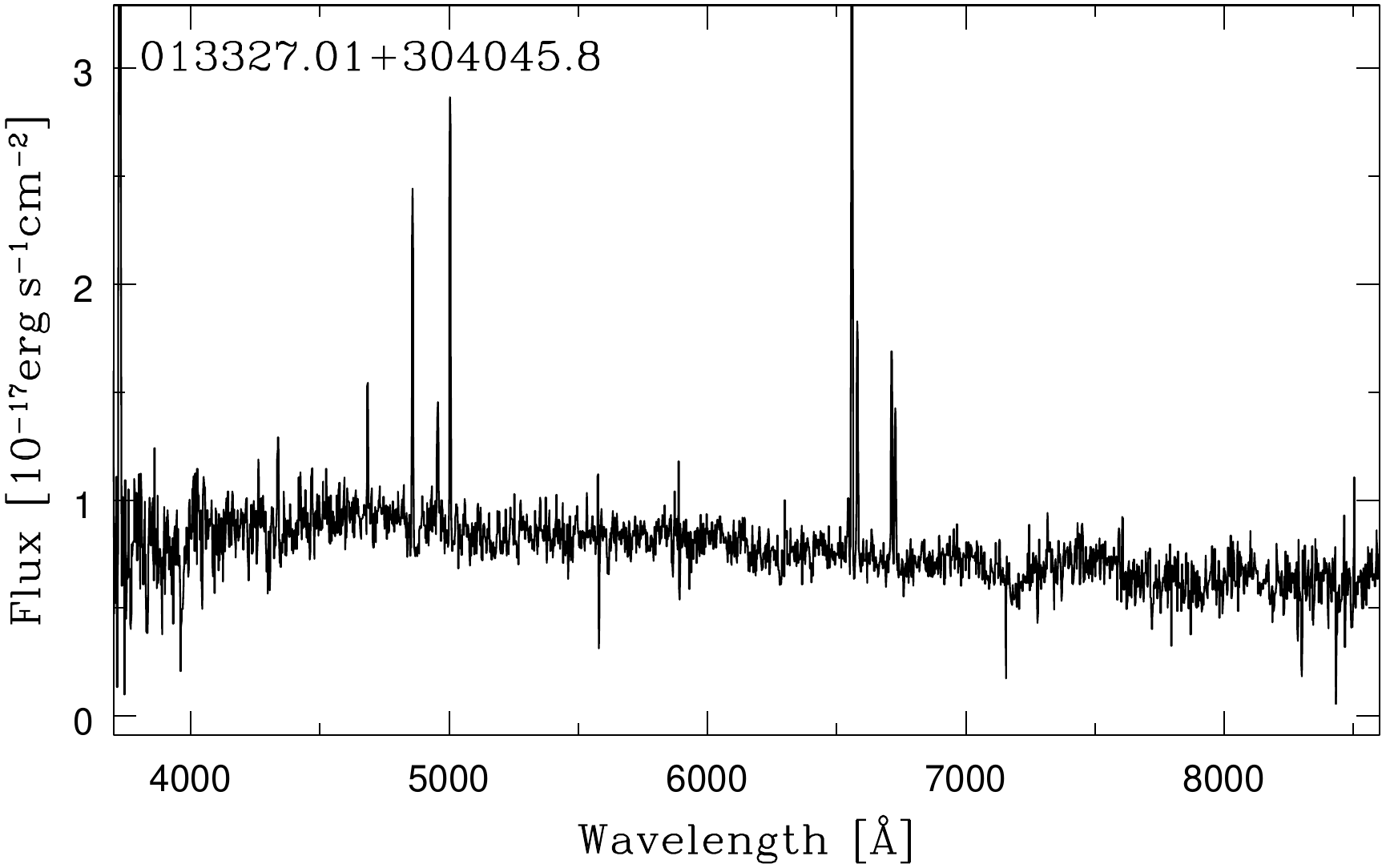}}
\centerline{\includegraphics[width=0.99\columnwidth]{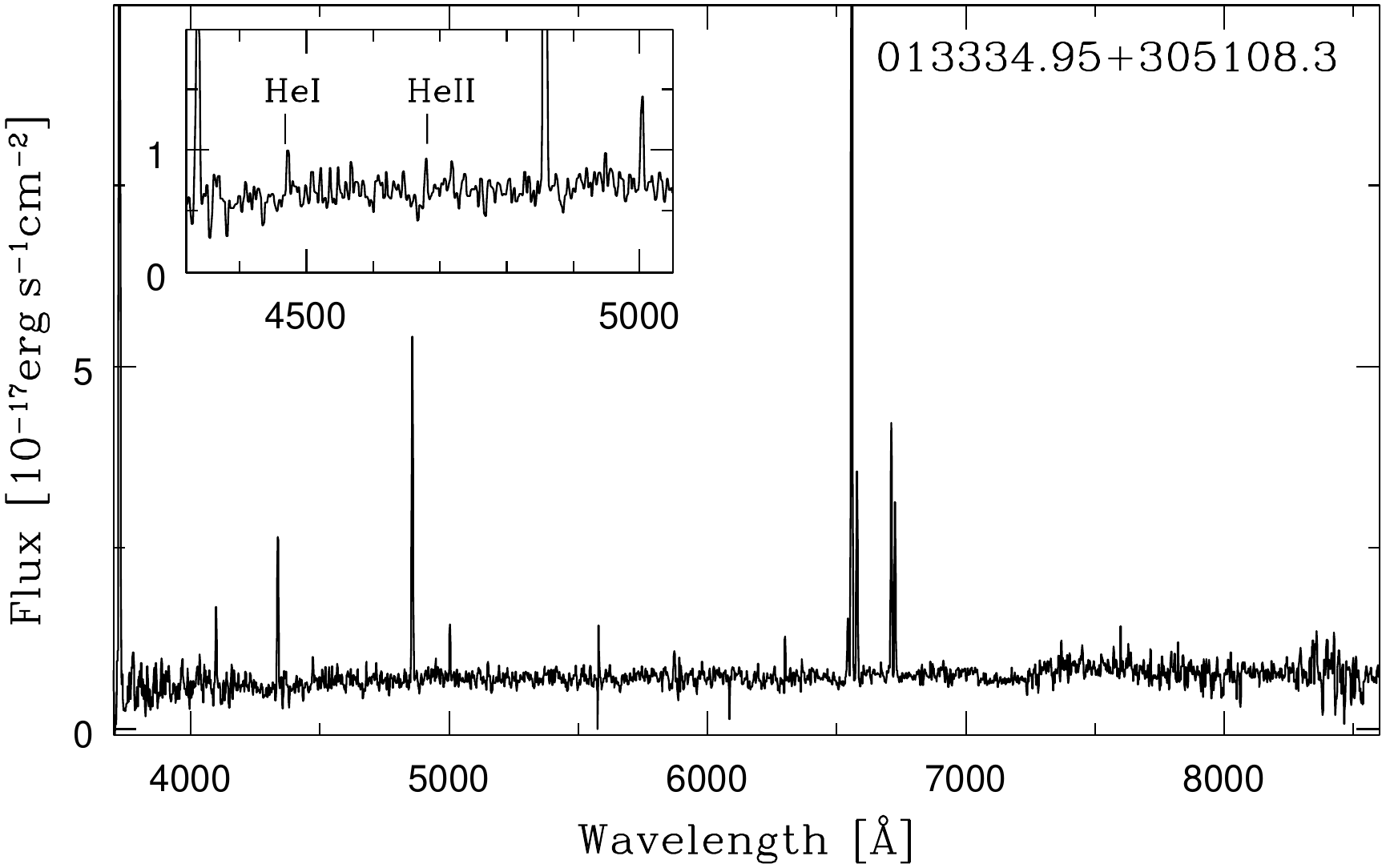}}
\centerline{\includegraphics[width=0.99\columnwidth]{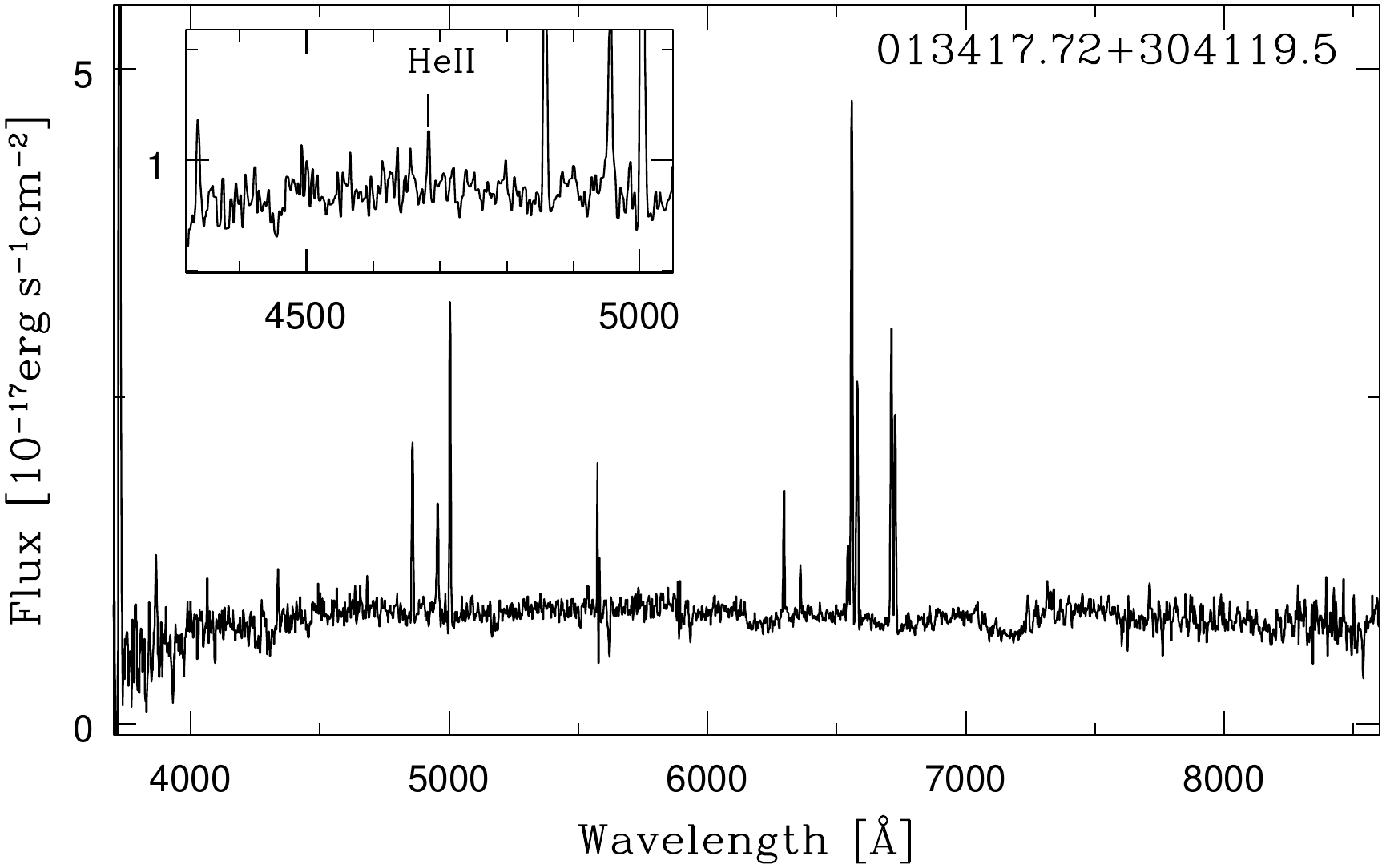}}
\caption{Spectra of SySt in M33 with some DIG emission.}\label{spectra3}
\end{figure}

\begin{figure}
\centerline{\includegraphics[width=0.99\columnwidth]{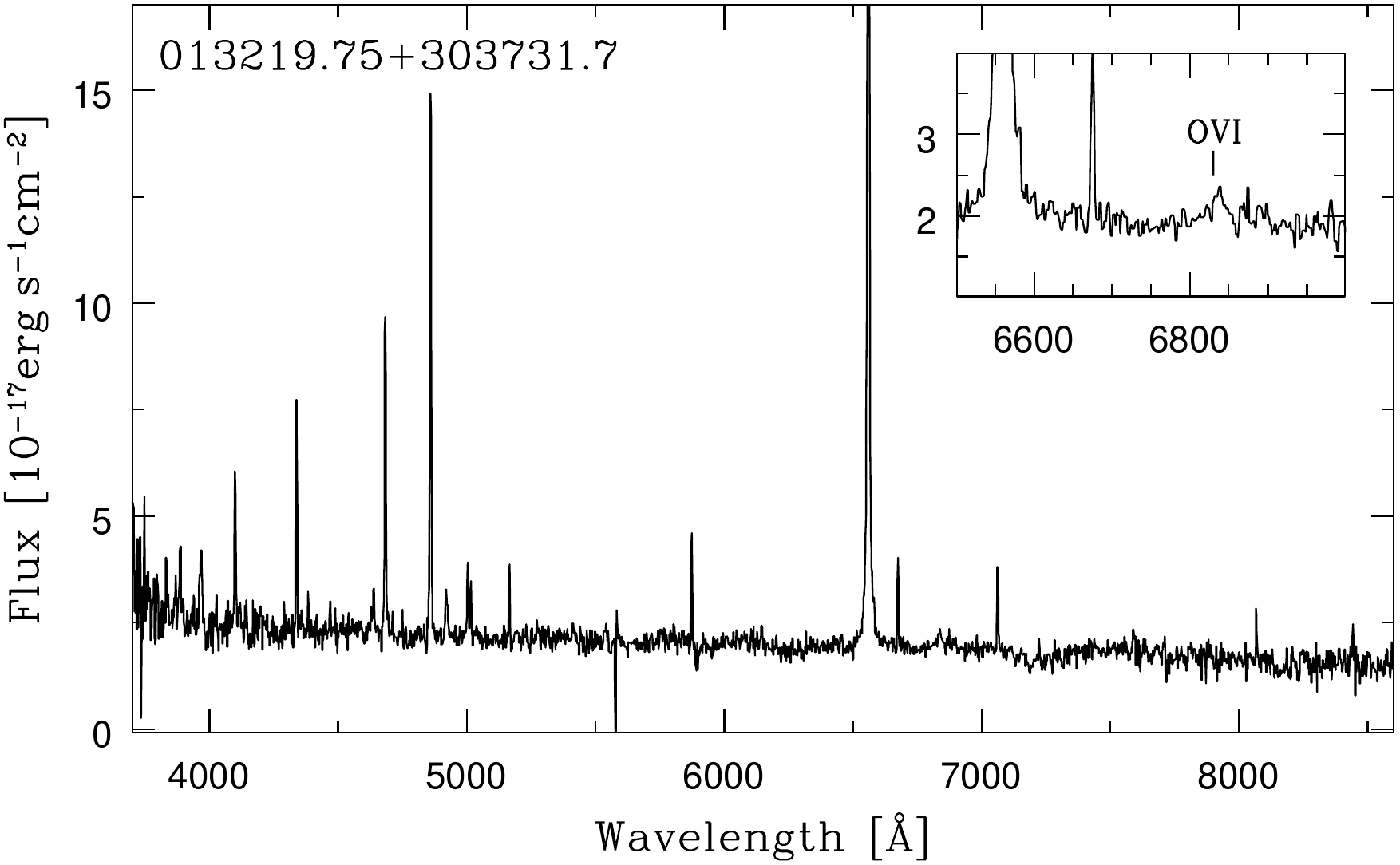}}
\caption{The spectrum of M33SyS\,J013219.75+303731.7 misclassified as a cepheid in \citet{hartman2006}. The red continuum and the presence of  \mbox{O\,{\sc vi}} in addition to strong  \mbox{He\,{\sc ii}}, however, demonstrate that this must be a symbiotic star. }\label{sp_359130}
\end{figure}

\begin{figure}
\centerline{\includegraphics[width=0.99\columnwidth]{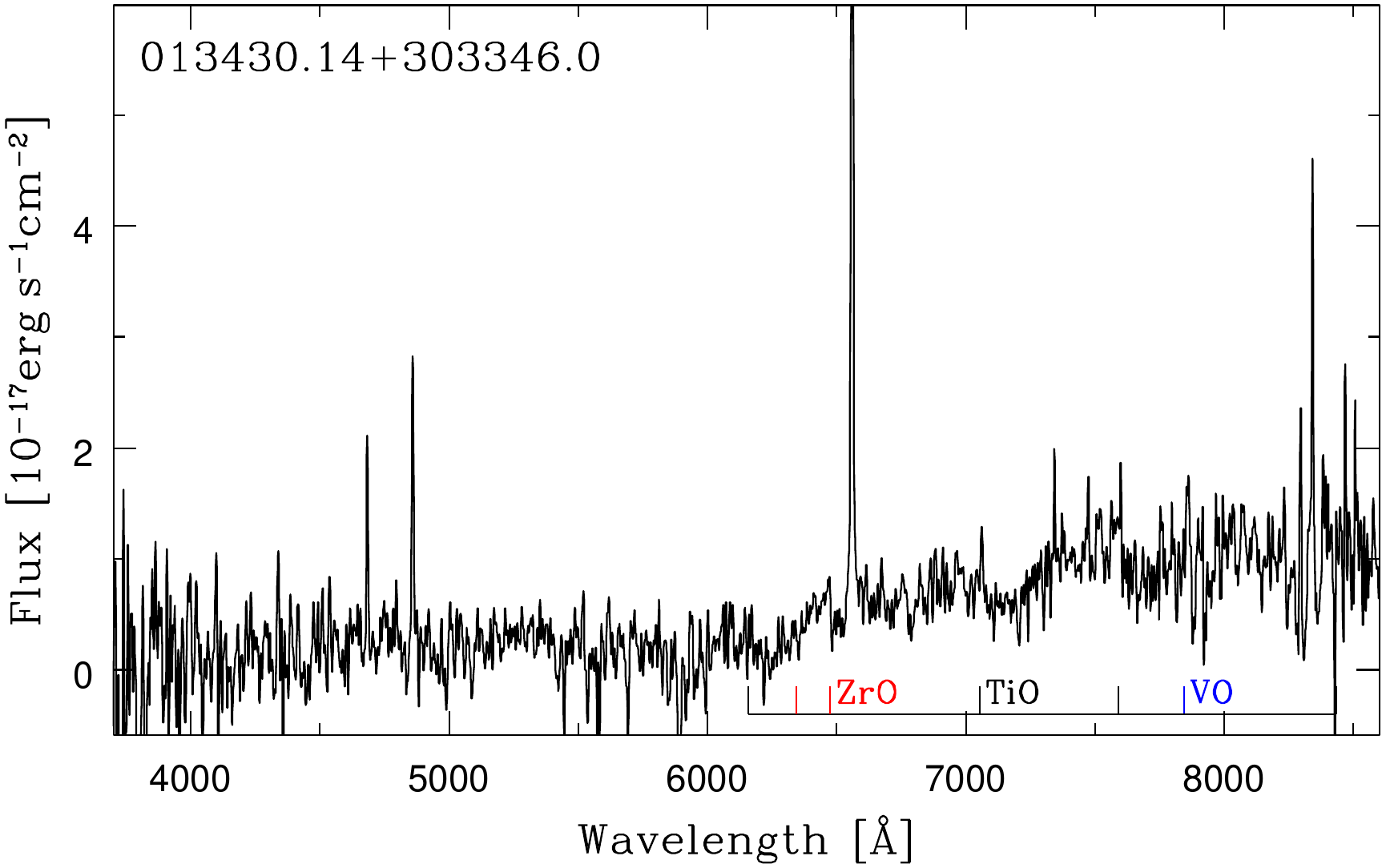}}
\caption{The spectrum of M33SyS\,J013430.14+303346.0 misclassified as a W Vir type variable in \citet{kinman1987}. The presence of strong red giant absorption features together with \mbox{He\,{\sc ii}} and \mbox{O\,{\sc vi}} emission lines indicates that this is a symbiotic star. }\label{sp_91161}
\end{figure}

All of our new SySt sample show  the \mbox{He\,{\sc ii}}\,$\lambda4686$ emission line and five of them also show the Raman-scattered \mbox{O\,{\sc vi}} lines. The \mbox{[Fe\,\sc{ vii}]}\,$\lambda6086$ line is visible in only 2 of our SySt, M33SyS\,J013457.79+310054.2 and M33SyS\,J013303.27+303528.3, which are also the hottest as indicated by their very high ratio of \mbox{He\,{\sc ii}}\,$\lambda4686$/H$\beta$ (Table~\ref{Tsp}).

The C-rich M33SyS\,J013435.17+303409.4 shows the faintest emission lines (Table~\ref{Tsp}, Fig.~\ref{spectraC}), and it does not show up as an H$\alpha$-bright object in the LGGS images (Fig.~\ref{fc2}). Thus, it should not have been selected according to our criteria, and the colour remeasured for this paper is H$\alpha - R \sim 0$ (Table~\ref{Tphot}). We suspect that there must be some error in our first measurement which gave  H$\alpha - R \sim $--3.7!
 
Finally, we note that three of our objects, M33SyS\,J013327.01+304045.8, M33SyS\,J013334.95+305108.3, and M33SyS\,J013417.72+304119.5, show relatively strong lines from the \mbox{[O\,\sc{ ii}]},  \mbox{[N\,\sc{ ii}]} and \mbox{[S\,\sc{ ii}]} lines (Fig.~\ref{spectra3}) which might suggest DIG origin. Although the low-density forbidden lines of \mbox{[N\,\sc{ ii}]} and \mbox{[O\,\sc{ ii}]} can be very strong in some SySt, especially those with a Mira-type giant (a very good example is \mbox{Hen\,2--147}, see figure  43 in \citealt{mz2002}), and symbiotic novae, they usually do not show strong \mbox{[S\,\sc{ ii}]}\,$\lambda6717,6731$ lines. 
The \mbox{[N\,\sc{ ii}]}/H$\alpha$, \mbox{[S\,\sc{ ii}]}/H$\alpha$ and  \mbox{[S\,\sc{ ii}]}(6717/6731) ratios locate these three objects near the borderline between \mbox{H\,\sc{ii}} regions and shock-dominated nebulae (e.g. \citealt{canto}, \citealt{pc1999}). 
Strong \mbox{[S\,\sc{ ii}]}\,$\lambda6717,6731$ lines were found in extended nebulae of Sanduleak's star, the jet--dominated  LMC SySt \citep{angeloni2012} and symbiotic--like \mbox{M\,2--9} \citep{pc1999}.
In M33SyS\,J013327.01+304045.8 the \mbox{He\,{\sc ii}}\,$\lambda4686$ is relatively strong, while in the remaning  two objects the line is relatively weak but detectable (Fig.~\ref{spectra3}). Moreover,  the position of all three in the \mbox{[O\,\sc{ iii}]} diagnostic diagram (\citealt{gmmc1995}; see also MCS14) indicate that they are SySt. 
Thus we maintain that they are likely SySt with some component of shock excitation and/or polluted by DIG which is very abundant in M33.

\section{Characterization of the new S\lowercase{y}S\lowercase{t} in M33}\label{analysis}

\begin{figure}
\centerline{\includegraphics[width=0.99\columnwidth]{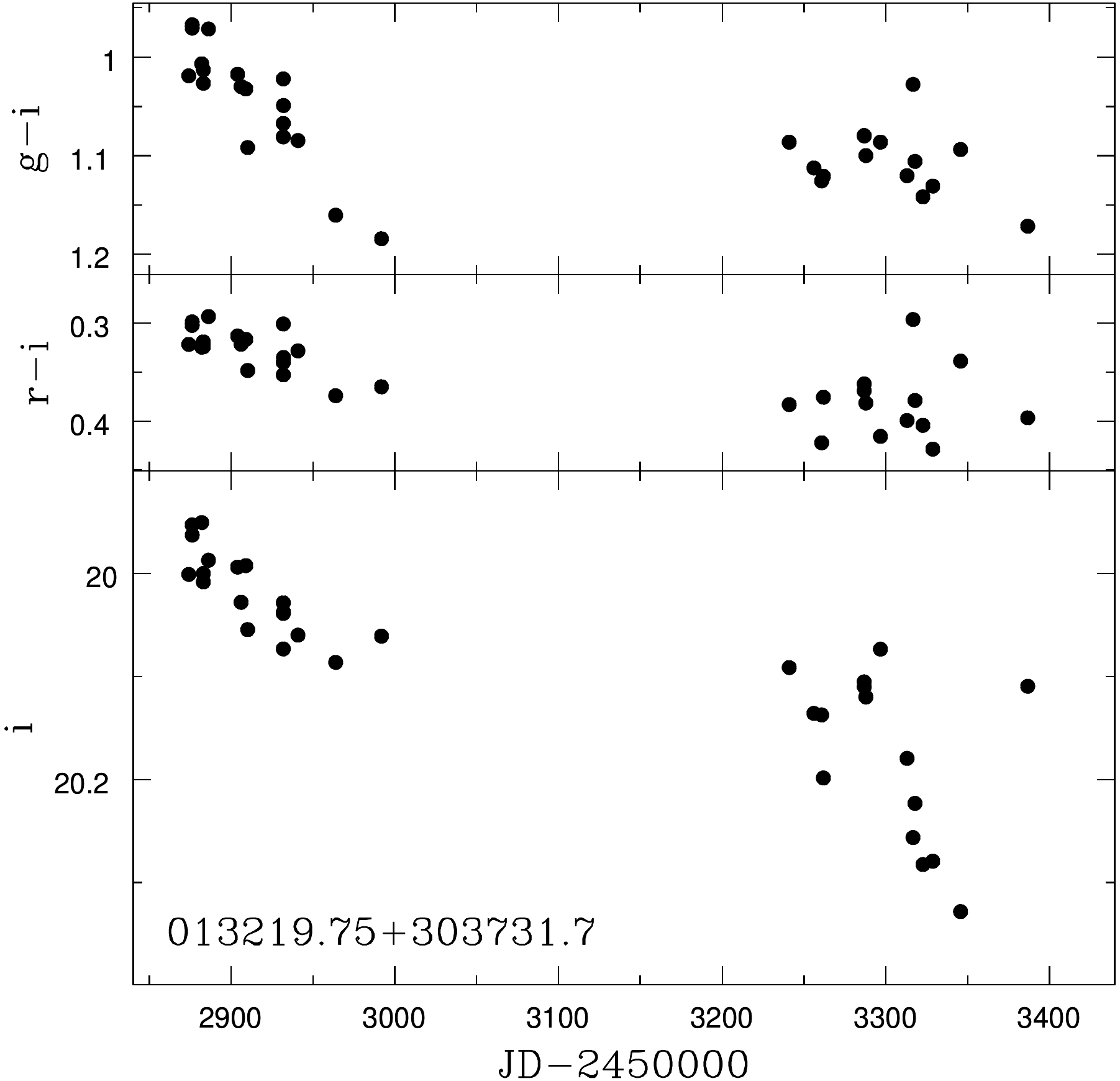}}
\caption{The CFHT light curve of M33SyS\,J013219.75+303731.7 misclassified as a cepheid in \citet{hartman2006}. }\label{lc_359130}
\end{figure}

\begin{figure}
\centerline{\includegraphics[width=0.99\columnwidth]{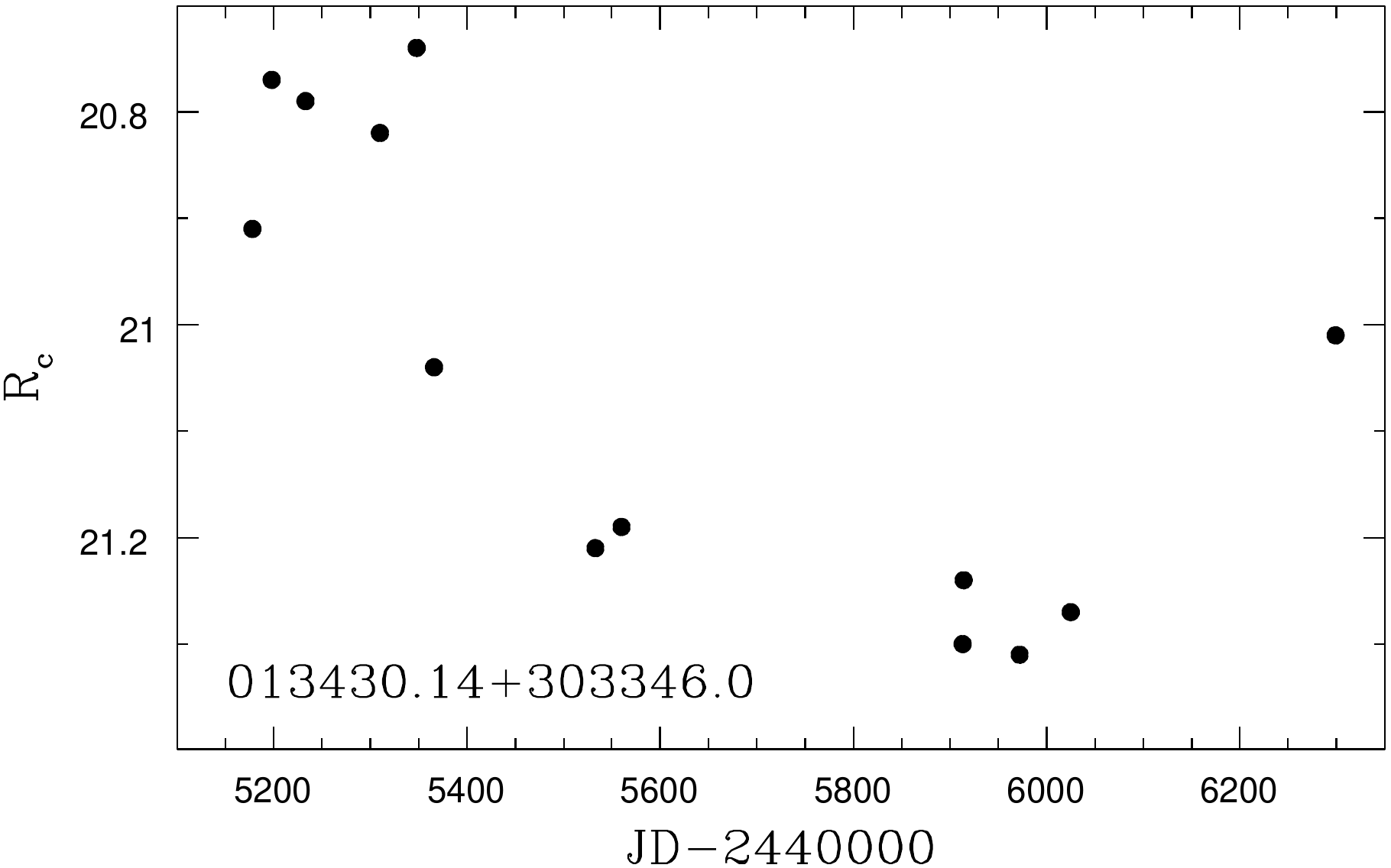}}
\caption{The light curve of M33SyS\,J013430.14+303346.0 misclassified as a W Vir type variable in \citet{kinman1987}. The presence of strong red giant absorption features together with \mbox{H\,{\sc ii}} and \mbox{O\,{\sc vi}} emission lines indicates that this is a symbiotic star. }\label{lc_91161}
\end{figure}

\subsection{Variability}\label{variability}

Light curves for 8 objects are available in the variability survey of M33 conducted with the 3.6-m Canada-France-Hawaii Telescope (CFHT) by \citet{hartman2006}. The data were obtained in 
$gri$ (Sloan filters), and the observations were made for 27 nights spanning 17 months. 

\begin{figure*}
\centerline{\includegraphics[width=0.99\columnwidth]{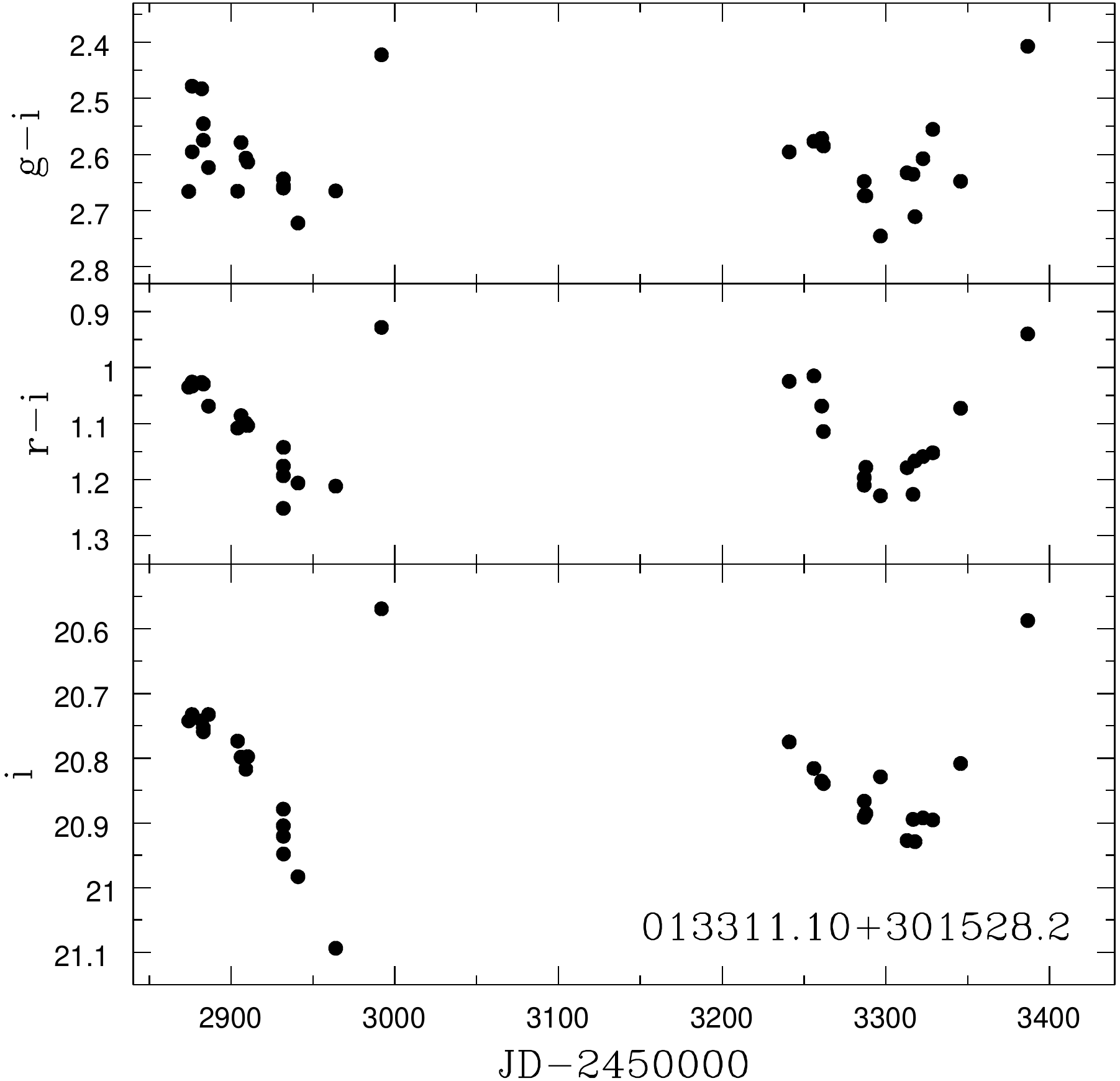}
\includegraphics[width=0.99\columnwidth]{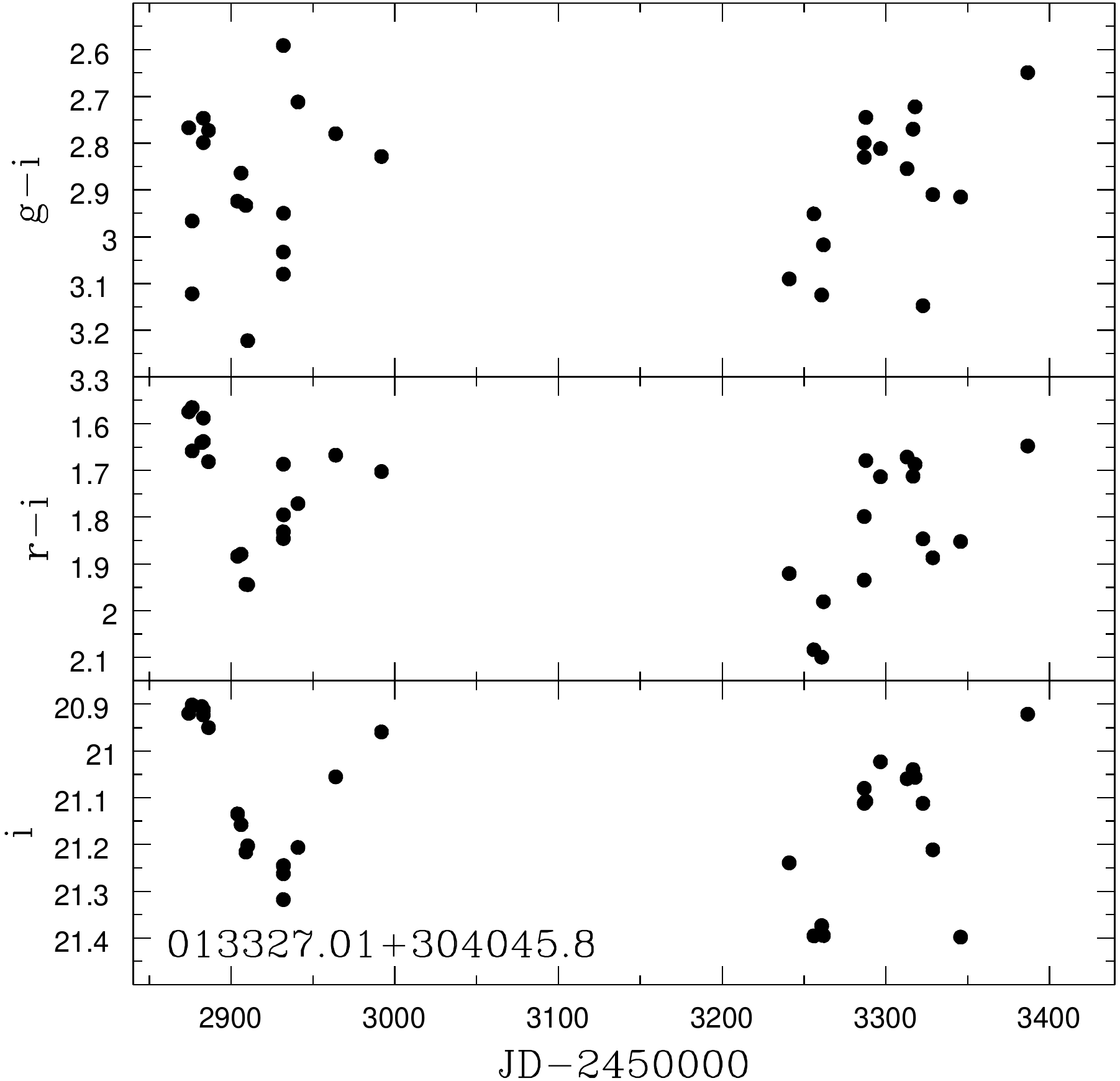}}
\centerline{\includegraphics[width=0.99\columnwidth]{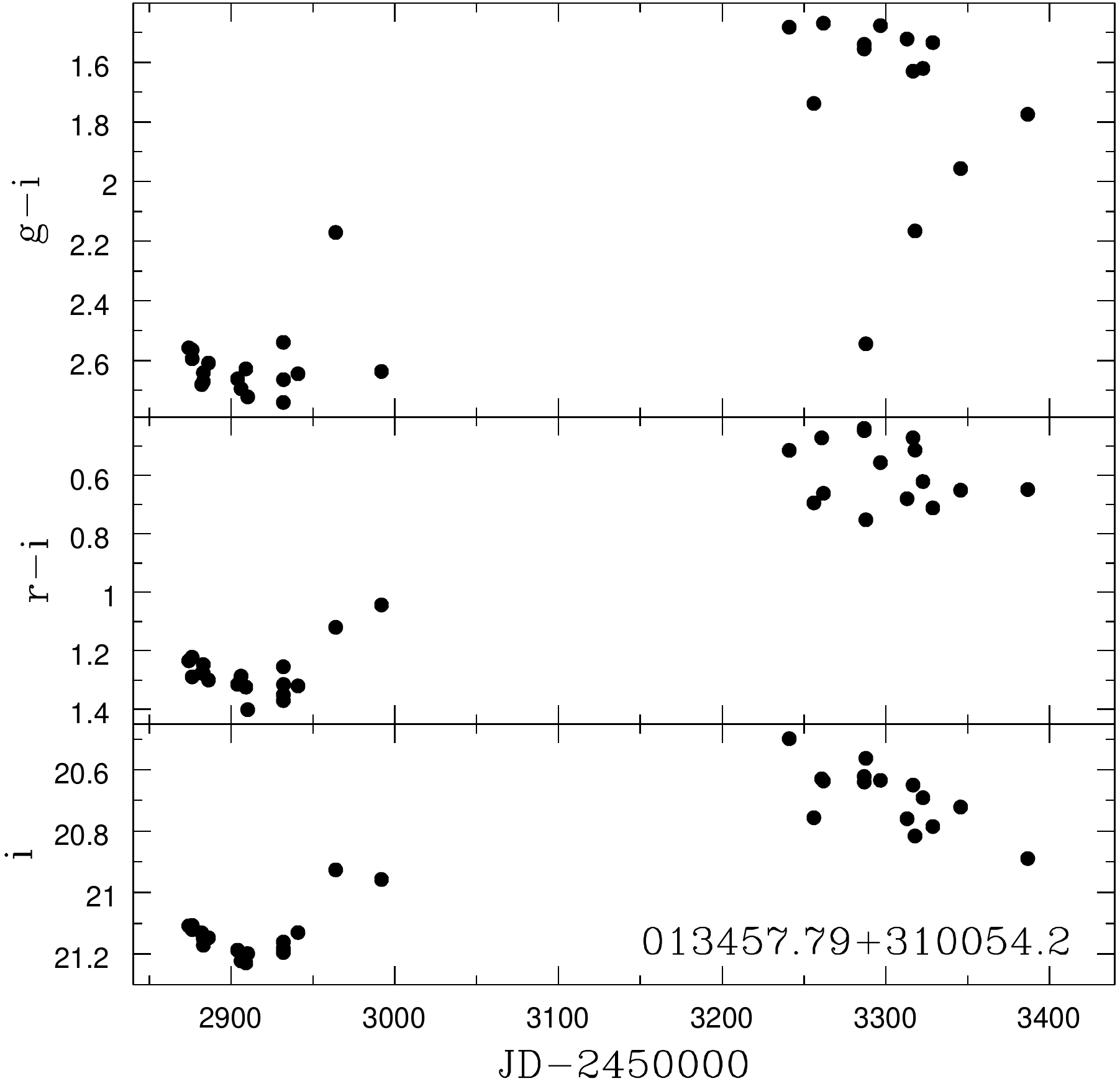}
\includegraphics[width=0.99\columnwidth]{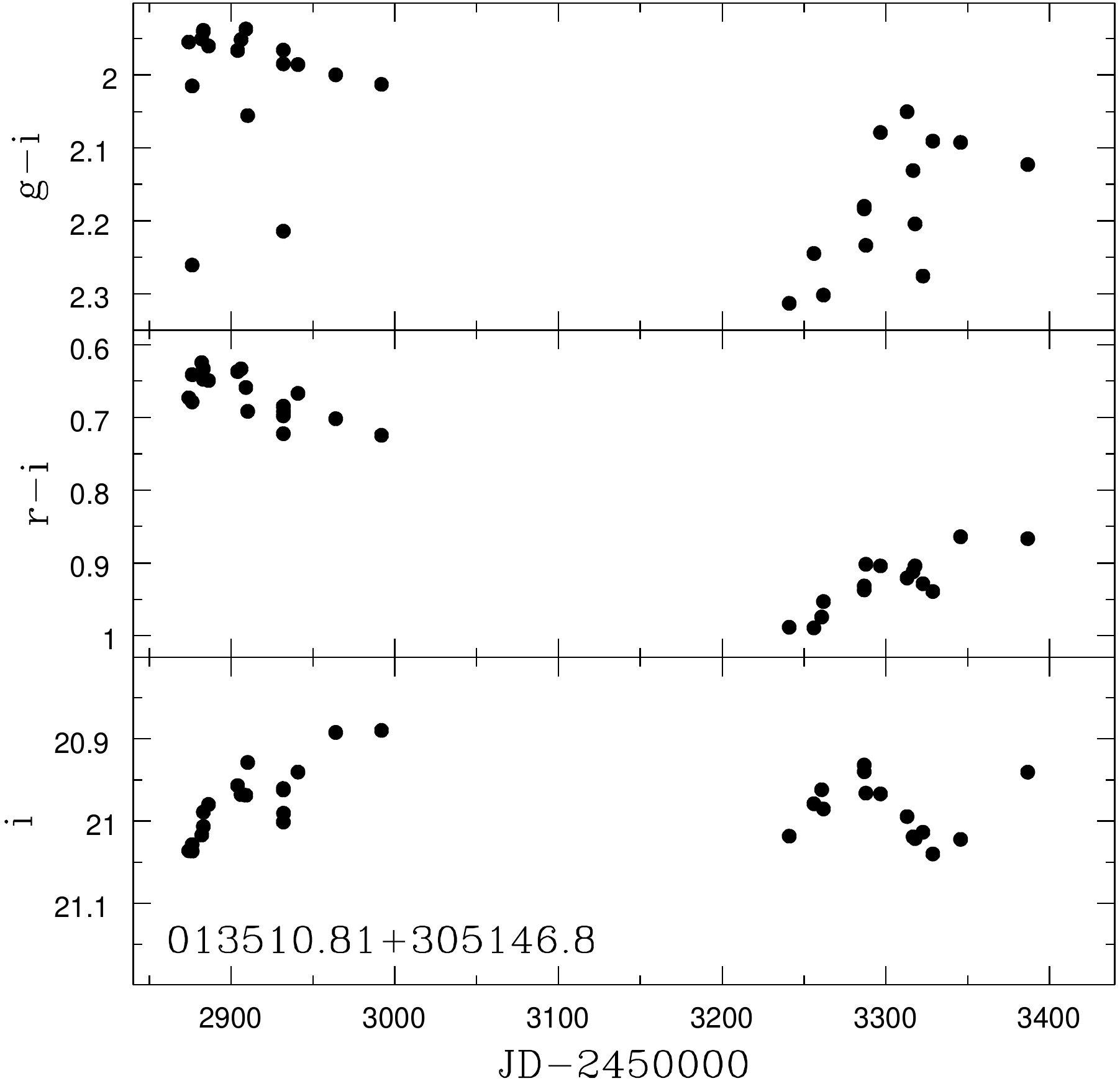}}
\caption{The CFHT light curves of SySt in M33 with M/MS giants.}\label{lc_M}
\end{figure*}

\begin{figure}
\centerline{\includegraphics[width=0.99\columnwidth]{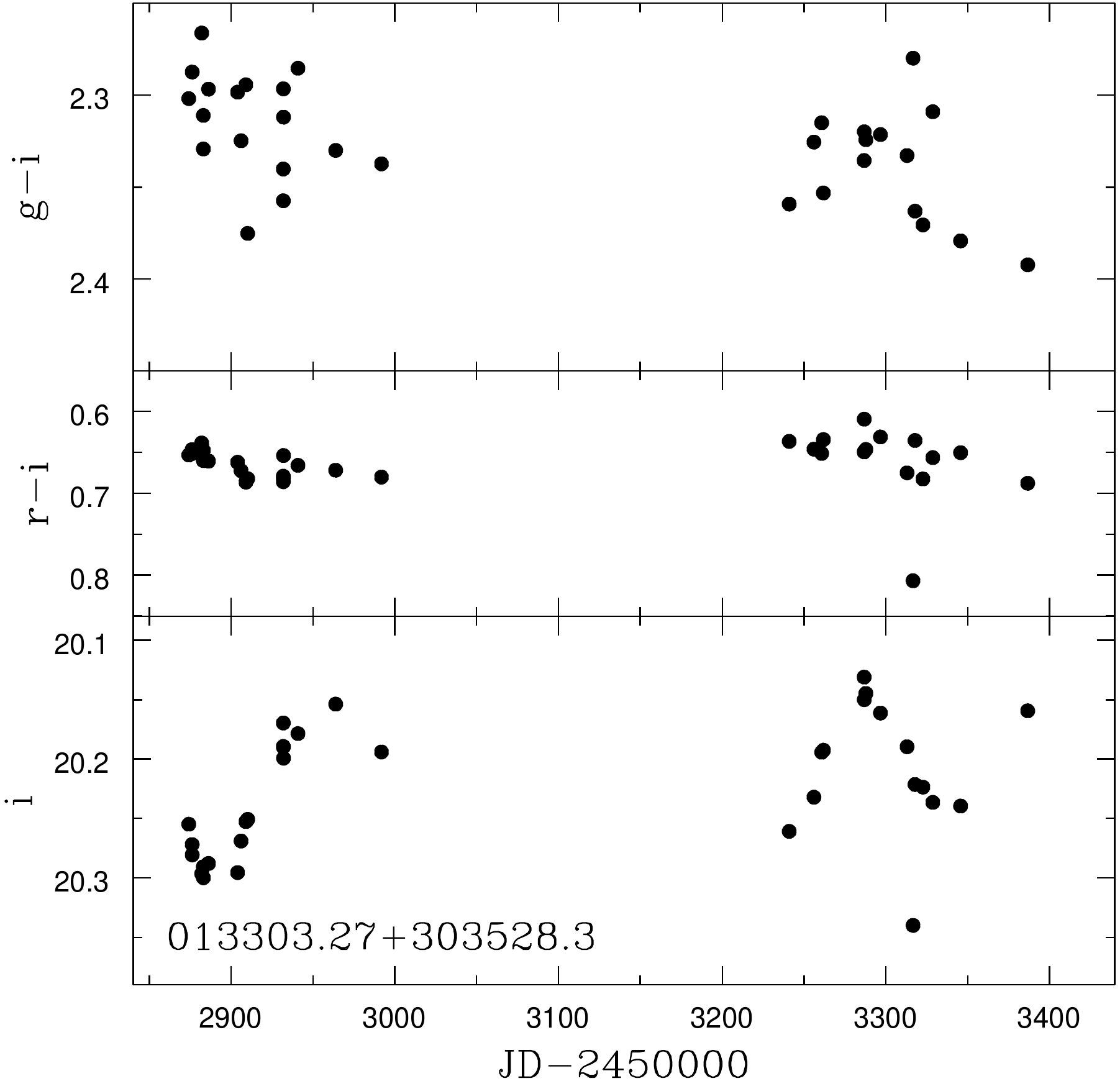}}
\centerline{\includegraphics[width=0.99\columnwidth]{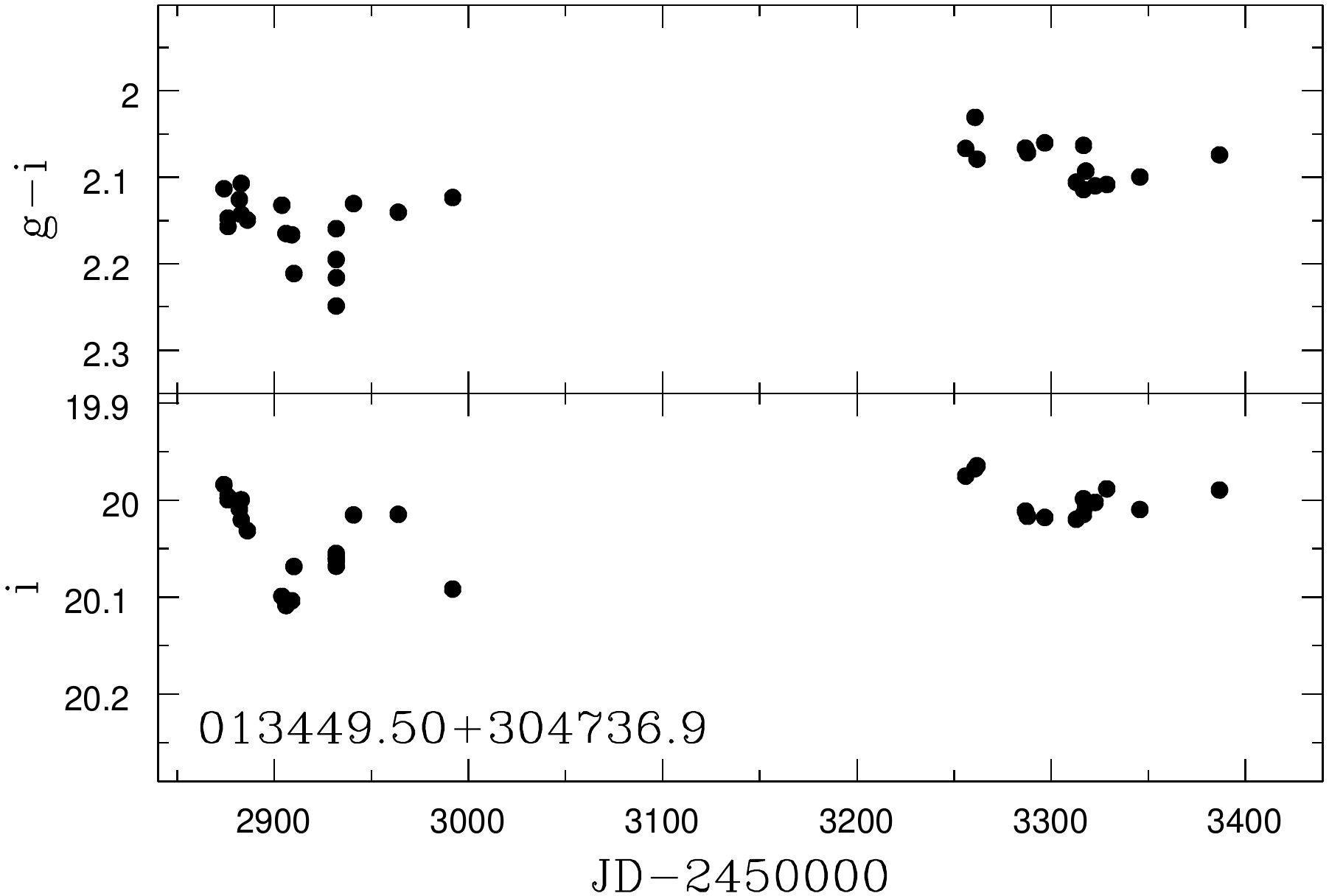}}
\centerline{\includegraphics[width=0.99\columnwidth]{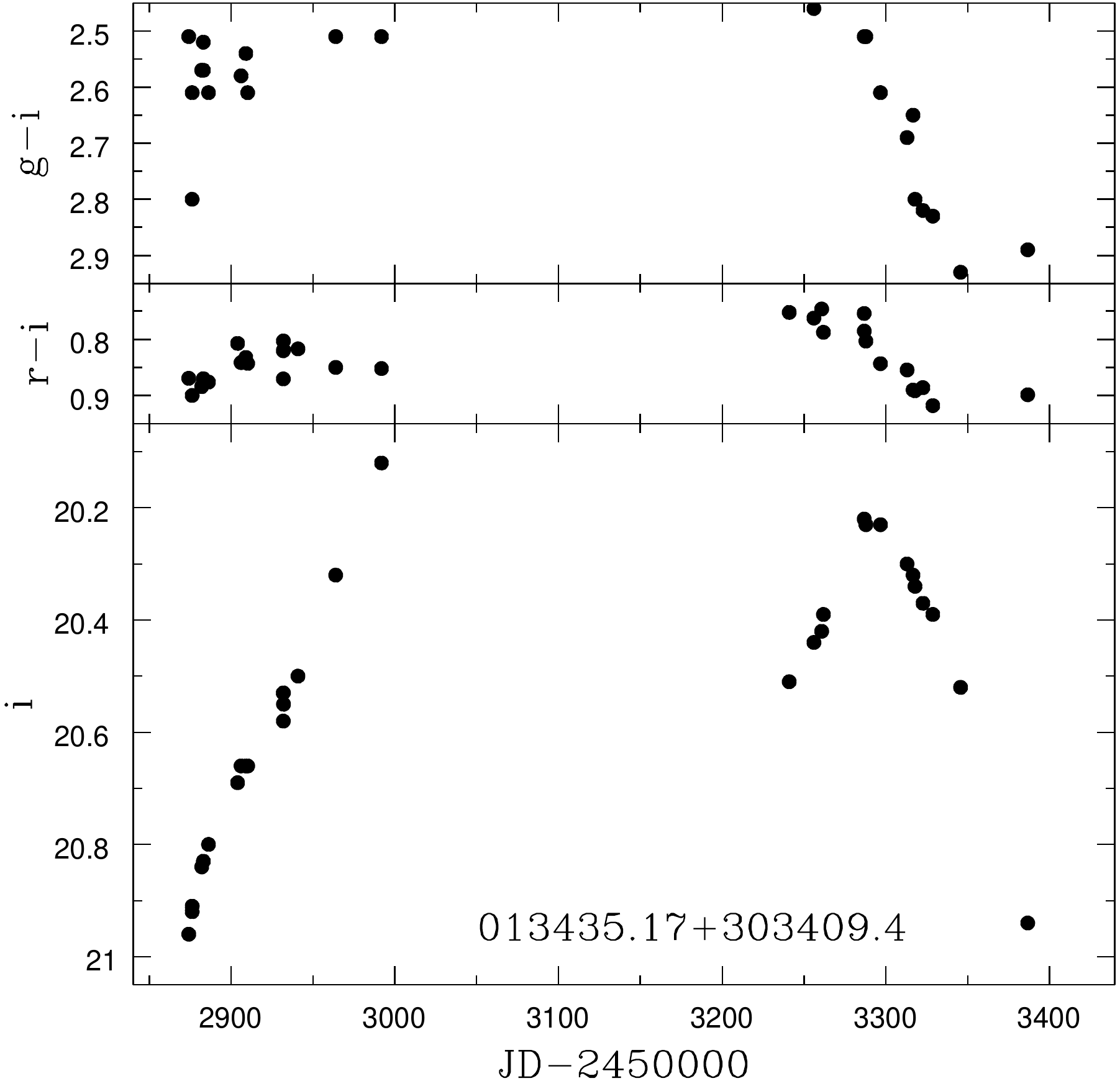}}
\caption{The CFHT light curves of SySt in M33 with C giants. }\label{lc_C}
\end{figure}

\citet{hartman2006} have classified all these objects as long-period variables, except for M33SyS\,J013219.75+303731.7 which was classified as a Cepheid with a 23.11 day period. 
However, both the light curves of M33SyS\,J013219.75+303731.7 phased with this period in \citet{hartman2006}  as well as its symbiotic  spectrum  (Fig.~\ref{sp_359130}) are totally incompatible with such a classification. Thus, we retrieved the original measurements, and the resulting light curves  shown in Fig.~\ref{lc_359130} are similar to those observed for the seven remaining  SySt (Figs.~\ref{lc_C}, \ref{lc_M}).
The reason for this misclassification seems to be that \citet{hartman2006} based their classification on magnitude-colour diagrams (their figure 7) which separate the Cepheid region from other classes of variable stars. 
Although the relatively blue $g-r$ and $r-i$ colours of M33SyS\,J013219.75+303731.7 are indeed consistent with a Cepheid, SySt with their composite spectra can easily mimic early spectral types. A very good example is the well-studied Galactic SySt AR Pav which shows the red giant colours only during eclipses. Outside the eclipse, its $V \sim 10.6$, $(V-R_{\rm C}) \sim 0.7$ and $(V-I_{\rm C}) \sim 1.4$ \citep{menzies82} which roughly correspond to $r\sim 21.4$, $g-r \sim 0.4$ and $r-i \sim 0.5$ (adopting the relations from \citealt{fukugita96}) will locate AR Pav in the Cepheid region rather than that occupied by M giants.

Similarly, M33SyS\,J013430.14+303346.0 was classified as a W Vir star with P=29.12 by \citet{kinman1987} who published only the phased light curve. 
Fortunately, we managed to recover the original light curve using the dates of observations (table I in \citealt{kinman1987}), and the resulting light curve is shown in Fig.~\ref{lc_91161}. It is clear that the average gaps  between the observations are comparable to the period found by \citet{kinman1987}, and the light curve is dominated by long-term changes. The  $B-R_{\rm C} \sim 1.7$ measured for M33SyS\,J013430.14+303346.0 is similar to those of other W Vir variables (table IV in \citealt{kinman1987}), however it is also significantly redder than the out-of-eclipse $B-R_{\rm C} \sim 1.3$ in AR Pav \citep{menzies82}.
Similarly, the out-of-eclipse $B-R_{\rm C} \sim 1.3$ in AR Pav is also bluer than $B-R_{\rm C} \sim 1.7$ measured for M33SyS\,J013430.14+303346.0 by \citet{kinman1987}.

The light curves of all our SySt (Fig.~\ref{lc_359130}--\ref{lc_C}) are similar to those observed for the Galactic and Magellanic SySt ( \citealt{marg2009}; \citealt{marg2013};  \citealt{angeloni2014}). Unfortunately, it is not easy to distinguish between the red giant pulsation and orbitally related variations because both became redder in $gri$ bands at minimum light.
However, the colour changes are larger for larger amplitudes and longer pulsation periods  (Mira or LPV variables), and often negligible in low-amplitude semiregular (SR) variables. 
The orbital variability can be detected only in S-type SySt where the red giant is nonvariable or an SR variable (\citealt{marg2013}; \citealt{ilkiew2015}).
In these SySt, the orbital modulation is best visible in the visual and blue bands while the low amplitude SR pulsations in these SySt dominate in the red bands (in SySt with strong emission lines the SR pulsations are visible only during the orbital minimum).  
In fact, the light curve and colour behaviour of M33SyS\,J013219.75+303731.7 is similar to orbitally related changes in S-type SySt, and rather does resemble the red giant pulsations.
The case of M33SyS\,J013430.14+303346.0 is less clear because there is no information about the colour changes.

The CFHT light  curves for the remaing O-rich SySt are shown in Fig.~\ref{lc_M}.
Among these, only M33SyS\,J013510.81+305146.8 shows clear indication that its varibility, especially in the $g,r$ bands is presumably due to an orbital motion with low-amplitude SR pulsation of the red giant dominating in the $i$ band. 
The light curves of M33SyS\,J013311.10+301528.2 and M33SyS\,J013327.01+304045.8 seem to be dominated by the red giant pulsations. 
The case of M33SyS\,J013457.79+310054.2 is less obvious, however, if the prominent light modulation is due to Mira-type pulsations, the period should be relatively long, $\ga 300^{\rm d}$. However, its observed $K$ (Table~\ref{Tphot}) corresponds to the absolute magnitude $M_{\rm K} \sim -6.7$ at the distance of M33 (see Section~\ref{par}), which is too faint for a Mira. In fact, all Galactic symbiotic Miras with similar periods have $M_{\rm K} \la -7.8$ \citep{marg2009}.
Thus we think that the variability of this SySt is due to orbital motion.

The CFHT light  curves of our C-rich SySt are shown in Fig.~\ref{lc_C}.
M33SyS\,J013435.17+303409.4 is definitely a Mira variable with a period of $\sim 300^{\rm d}$. The observed $K$ magnitude (Table~\ref{Tphot}) corresponds to the absolute magnitude $M_{\rm K} \sim -7.9$, which is also consistent with a Mira variable and $P\sim 300^{\rm d}$ \citep{marg2009}. Its spectrum (Fig.~\ref{spectraC}) is very similar to the symbiotic C-rich Mira H1-45 \citep{MMU2013}.
It is also the coolest SySt in our sample, as indicated by its optical and $JHK$ magnitudes and colours. However its $JHK$ magnitudes and colours do not show any evidence for an optically thick dust found in the majority of Galactic symbiotic Miras.

The light curves of M33SyS\,J013303.27+303528.3 also seem to be dominated by the red giant pulsation, however, with a significantly lower amplitude and a shorter period ($\sim 100^{\rm d}$). 
Finally, in M33SyS\,J013449.50+304736.9, the $g-i$ colour is redder when the low-amplitude ($\Delta\,I \la 0.1$) oscillations are visible, and it became bluer when the star became brigher (mean $I$ magnitude is lower), and the oscillations disappear. Thus, its 
light curve can be a combination of the low-amplitude SR pulsations of the giant and orbitally related changes with a longer ($\ga 500^{\rm d}$) period. Similar behaviour is observed  in the C-rich SySt LMC S63 \citep{ilkiew2015}.

\subsection{Physical parameters}\label{par}

The physical parameters of our symbiotic sample are presented in Table~\ref{Tpar}. 
The hot component temperature, $T_{\rm h}$, was estimated in the same way as by MCS14. In particular, the lower limit was set by the highest ionization stage, $IP_{\rm max}$, using the formula 
$T_{\rm h}$[1000\,K]=$IP_{\rm max}$[eV] \citep{MN94}. The upper limit for $T_{\rm h}$ was set by the \mbox{He\,{\sc ii}}4686,  \mbox{He\,{\sc i}}5876, and H$\beta$ line ratios assuming case B recombination and cosmic He/H \citep{iijima1981}. 

The foreground extinction towards M33 is very low, $E(B-V) \sim 0.03$  \cite{schlafly2011}, and the Balmer line ratios H$\alpha$/H$\beta$/H$\gamma$ are not very useful for reddening estimates in SySt because of their significant departures from case B conditions, presumably due to self absorption effects (e.g. \citealt{proga1996}, and discussion in MCS14). Thus, the absolute $I$ magnitudes, $M_{\rm I}$ and the total H$\alpha$ luminosities, $L(\rm H\alpha)$, were calculated assuming  $E(B-V) \sim 0$, and adopting the the true distance modulus m--M=24.62 ($d$=835 kpc) \cite{gie13}. 
The comparison of the H$\alpha$ fluxes measured on our spectra with those derived from the H$\alpha$ magnitudes, and listed in Table~\ref{Tsp} shows that they agree within a factor of $\sim 2$, and such a discrepancy can be easily accounted for by their intrinsic variability (e.g. \citealt{ilkiew2015}; see also Section~\ref{variability}). To calculate $L(\rm H\alpha)$ we used the H$\alpha$ fluxes derived from spectra; they are accurate within a factor of $\sim 2$.

As in the case of SySt in M31 and the Milky Way, there is some positive correlation between  $M_{\rm I}$ and $L(\rm H\alpha)$. The H$\alpha$ luminosity characterizes the nebular emission which is related to the hot component temperature and luminosity while $M_{\rm I}$ characterizes the cool giant. Thus, this correlation probably reflects a relation between the hot component luminosity and that of the cool giant. 

\begin{table*}
 \centering
  \caption{Emission line ratios (H$\beta$=100) measured in the spectra of the new M33 SySt. In addition, the last column gives  the H$_\alpha$ fluxes measured on the spectra (first value) and those derived from the H$\alpha$ magnitudes (second value).}\label{Tsp}
  \begin{tabular}{@{}lrrrrrrrrrrrcrrc@{}}
\hline
 M33SyS\,J  & [O\,{\sc ii}]$^1$ & [Ne\,{\sc iii}] & H$\gamma$ & [O\,{\sc iii}] & He\,\sc{ii} & [O\,{\sc iii}] & He\,{\sc i} &  [Fe\,{\sc vii}] & H$\alpha$ & [N\,{\sc ii}] & He\,{\sc i} & [S\,{\sc ii}] & O\,{\sc vi} & He\,{\sc i}&
 $F$(H$\alpha$)$^1$ \cr
         &  3728  &  3868 &   & 4363&  4686&  5007 & 5876 & 6086 &  & 6583 & 6678 & 6716/31 & 6825 & 7065  & \cr
\hline
013219.75+303731.7 & & &  38 & 8: & 61 & 13  & 22 & & 503 &  & 16 & & 2 & 16 &   41/23 \cr
013239.11+303836.5 & & & &  & 51 & 108 & 66 & & 445 & & & & &  & 3/5\cr
013303.27+303528.3 & 205 & &  40 & 9: & 97 & 45 & 29 & 11 & 575 &  30 & 10 & 35/25 & 35 & 24  & 16/14 \cr
013311.10+301528.2 &  & 100 & & & 72 & & 26 & &  523 & & & & 17 &  & 25/10 \cr
013327.01+304045.8 & 536 &  & 40 & 17 & 40 & 147 & 25 & & 336 & 76 & & 67/47 & & 13  & 3.3/3.6 \cr
013334.95+305108.3 & 365 & 21 &  52 & 27 & 10 & 17 & 12 & & 335 & 65 & 6 & 77/53 & &  & 9/6 \cr
013417.72+304119.5 & 865 & 81 & 41 & 13 & 21 & 193 & &  & 322 & 167 & & 186/136 & & &  3/6\cr
013430.14+303346.0 & & 37 & 25 & 24 & 76 & 19 & 12 & &  480 & &16 & & 10 & 29 &  12/9 \cr
013435.17+303409.4 &  & & & & 43 & & & & 374 & & & & &  & 1.4/0: \cr
013449.50+304736.9 & & &   43 & & 48 & 22 & 13 & & 420 & & 11 &  & & 21 & 17/17 \cr
013457.79+310054.2 & & & 49 & 22 & 99 & 23 & 31 & 33 & 416 & & & & 40 & &  5/4\cr
013510.81+305146.8 & & 62 & 30 & 44 & 34 & & 21 & & 409 & & 32 &  & & 25 &  24/14 \cr
\hline
\end{tabular}
\begin{list}{}{}
\item $^{1}$ blend of 2 lines
\item $^{2}$ in units of $10^{-16}\, \rm ergs\,s^{-1}\,cm^{-2}$. 
\end{list}
\end{table*}

\begin{table*}
 \centering
  \caption{Physical parameters of new SySt in M33.}\label{Tpar}
  \begin{tabular}{@{}lccrcccc@{}}
\hline
 M33SyS\,J &  Sp Type &  Ion & $T_{\rm h}\,[10^3$\,K] & $M_{\rm I}$ & $L(\rm H\alpha)^1$  & $v_{\rm r}$\,[km\,s$^{-1}$] & $\Delta v\,[\rm km\,s^{-1}]^2$ \cr
\hline
 013219.75+303731.7 & K/M &  O$^{+5}$ & 114$\div$153 & -5.0 & 3.73/97 & -145$\pm$3 & 5 \cr
 013239.11+303836.5 & CH & He$^{+2}$ & 54$\div$132 & -3.9 & 0.27/7.2 &  -209$\pm$10 & -52\cr
 013303.27+303528.3 & CN & O$^{+5}$ & 114$\div$176 & -5.1 & 1.41/37& -169$\pm$5 & -12 \cr
013311.10+301528.2 & M5 & O$^{+5}$ & 114$\div$159 & -4.4 & 2.22/59 & -115$\pm$20 & -15 \cr
 013327.01+304045.8 & M & He$^{+2}$ & 54$\div$131& -4.1 &  0.30/8.3 &-170$\pm$7 & -5 \cr
 013334.95+305108.3 & M &  He$^{+2}$ & 54$\div$94 & -4.2 & 0.86/23 & -223$\pm$3 & 9 \cr
013417.72+304119.5  & M &  He$^{+2}$ & 54$\div$116 & -4.4 & 0.27/7 & -206$\pm$8  & 12 \cr
013430.14+303346.0 & M2S & O$^{+5}$ & 114$\div$169 & -4.5 & 1.12/29 & -166$\pm$13  & -1 \cr
 013435.17+303409.4 & CN & He$^{+2}$ & 54$\div$142 & -4.1 & 0.13/3.3 & -157$\pm$6 & 23 \cr
013449.50+304736.9 & CN & He$^{+2}$ & 54$\div$144& -5.2  & 1.56/41 &  -245$\pm12$ & -5 \cr
 013457.79+310054.2 & M4 & O$^{+5}$ & 114$\div$176 & -4.1 & 0.47/12&  -230$\pm$18 & 32 \cr
013510.81+305146.8 & M4S & He$^{+2}$ & 54$\div$127& -4.4 & 2.17/57 & -205$\pm$8 & 30 \cr
\hline
\end{tabular}
\begin{list}{}{}
\item $^{1}$ in units of $10^{35}\,\rm erg\,s^{-1}$/L$\sun$ 
\item $^{2}$ difference between the measured radial velocity $v_{\rm r}$ and the galactic rotational velocity \citep{m33rotation}
\end{list}
\end{table*}

The absolute magnitudes, $M_{\rm I}$, of the M33 SySt are between --3.9 and --5.2, and they cover more or less the same range as those of SySt in M31 (MCS14). They are all close to or above the tip of the Red Giant Branch (TRGB) in the colour-magnitude diagrams of resolved galaxies. In particular, \citet{udalski2000} determined the TRGB $I$-band magnitude, $M_{\rm I} \sim -4$, in the Magellanic Clouds. These absolute magnitudes are also consistent with the semi-regular character of CFHT light curves of most of our sample.

There is, however, a significant difference between the M33 and M31 samples in that 4 of the symbiotic giants in M33 are C-rich giants, and another 2 are MS stars whereas the M31 SySt all have O-rich, M-type giants (see table 4 in MCS14). C-rich SySt are also very rare in the Milky Way (e.g. \citealt{Bel2000}; \citealt{MMU2014}; \citealt{MM2014}).
The ratio of C to M giants is a strong function of metallicity of the parent galaxy which is low in M33, [M/H]$\la$--1.6 \citep{cioni2008}, and roughly solar in M31. 
The O-rich symbiotic giants in M33 are also on the average of earlier spectral type than those in M31. The C:M ratio of non-symbiotic giants found by our survey, C:M $\sim$ 0.3 is somewhat lower than that for SySt, however, this could be due to low number statistics. A similar effect is observed in the Magellanic SySt, where the ratio of C-rich to O-rich objects is 4:2 in the LMC, and 2:6 in the SMC, respectively, and most of the O-rich giants are of K or early M-type (\citealt{murset1996}; \citealt{MMU2014}).

\citet{murset1996} suggested that the Magellanic symbiotic giants are not influenced by binarity, and in particular, there is no need to account for the C abundance by a former mass transfer. However, this may be not true in the case of LMC S63 hosting a CJ-type giant \citep{murset1996}, and LMC N19 in which the M giant shows strong evidence for an s-process enhancement (Miko{\l}ajewska et al. 2016, in preparation). In both cases, the chemical abundances cannot be explained by a single star evolution.
Among the M33 SySt, M33SyS\,J013239.11+303836.5 contains a CH-type giant, similar to the CH-type metal-poor star component forming the majority of the Galactic halo carbon star population, and which have been ascribed to mass transfer in binaries  (\citealt{totten1998}, and references therein). Among the Galactic SySt, only two, both in the Galactic halo, are of CH-type, and both show strong evidence for carbon and s-process pollution by a former mass transfer \citep{schmid1994}.

\subsection{Distribution and ages of M33 S\lowercase{y}S\lowercase{t}}\label{distribution} 

Fig.~\ref{map}, which displays the locations of the 12 SySt we have identified in M33, demonstrates that these stars are not concentrated in the galaxy's nuclear regions, spiral arms, or star clusters. Exactly the opposite is true; the M33 SySt appear to be distributed ``randomly" across M33. 
A similar distribution is observed for SySt in M31 which are distributed over the whole disc up to $\sim 25\, \rm kpc$ from the galactic center \citep{mik2015}.
This sort of spatial distribution is a hallmark of old stellar populations, whose stars have orbited a galaxy multiple times and migrated far from their birthplaces.

Not a single SySt is a known member of a Galactic open cluster (OC) or globular cluster (GC). (This is not very surprising, as only $\sim 300$ Galactic SySt are known, and only a few percent of all Galactic stars are currently in clusters.) The result of this absence of cluster SySt is that the gold standard for age determination for any given Galactic SySt does not exist. 

We can address the question of the ages and age distribution of the M33 SySt - and hence those of all SySt - with the data in Table~\ref{Tpar}. \citet{beasley2015} have measured the radial velocities and ages of 77 star clusters in M33 to determine that galaxy's age-velocity relation (AVR), shown in their figure 19. 
The largest observed cluster velocity dispersion is 60 km\,s$^{-1}$. Most clusters' dispersions are in the 10--30 km\,s$^{-1}$ range, similar to those of most of our SySt. We conclude that the SySt of M33 are a disc population.

The only exception is M33SyS\,J013239.11+303836.5, with the velocity dispersion --52 km\,s$^{-1}$, and the only one with a CH-type giant. 
Table~\ref{Tpar} also shows that it is the least luminous star in our sample, displaying $M_{\rm I}$ = --3.9, somewhat below the TRGB luminosity (see Section~\ref{par}). Thus, this star must have mass $\la1.0 \,\rm M\sun$ \citep{schaller1992}, and hence age $\ga 10$ Gyr. 
The other SySt listed in Table~\ref{Tpar} are all more luminous than M33SyS\,J013239.11+303836.5, and hence more massive -- up to about 2\,M$\sun$ -- and with ages in the 1--10 Gyr range.   
We conclude that SySt can form at all times, and in all environments in M33 where binary stars are born. 

\citet{cioni2008} discussed the spatial distribution of C/M ratio, mean age, and metallicity of AGB stars in M33, and found that the average outer ring and nuclear stellar population is $\sim 6$ Gyr whereas the central regions are a few Gyr younger.
Our results are consistent with their study, as all of our C-rich SySt are located in the outer ring, and our oldest M33SyS\,J013239.11+303836.5 lies in the middle of the region occupied by the oldest ($\sim$\,10 Gyr) stars with the lowest metallicity (cf. figure 15 of \citealt{cioni2008}).


\section{Conclusions}\label{concl}

We presented and discussed initial selection criteria and first results in M33 from a systematic survey for extragalactic symbiotic stars. We found that most of the RGs with strong nebular emission are not SySt. It is DIG that is making these objects stand out as candidates, but they are not genuine SySt. Eliminating these imposters is essential if we are to isolate uncontaminated samples of genuine extragalactic SySt. 

We also presented the first 12 SySt detected in M33, and discussed their spectroscopic characteristics, and light curves (available for nine of these objects).
We found a high number ratio of C to M giants in these systems which is presumably due to the low metallicity of M33. 
The light curves of most of our SySt reveal semi-regular variability, similar to that observed in Galactic S-type SySt, which can be accounted for by a superposition of semi-regular pulsations of the red giant with $P \sim 100^{\rm d}$, and orbitally-related changes with a longer period.  
The only SySt hosting a Mira variable is the C-rich M33SyS\,J013435.17+303409.4, however its $JHK$ magnitudes and colours do not show any evidence for an optically thick dust found in the majority of Galactic symbiotic Miras.

The spatial and radial velocity distributions of the 12 new SySt we have identified in this paper are consistent with a wide range of progenitor star ages.

\section{Acknowledgments}

We gratefully acknowledge the fine support at the MMT Observatory, and
The Local Group Galaxy Survey conducted at NOAO by Phil Massey and
collaborators. This research has made use of the VizieR catalogue access tool, operated at CDS, Strasbourg, France.
This study has been supported in part by the Polish National Science Centre grant DEC-2013/10/M/ST9/00086. KI has been also financed by the Polish Ministry of Science and Higher Education Diamond Grant Programme via grant 0136/DIA/2014/43.


\newpage

\label{lastpage}

\end{document}